\documentclass[sigconf,nonacm]{acmart}

\AtBeginDocument{%
  }

\usepackage{xspace}
\usepackage{booktabs}
\usepackage{tcolorbox}
\usepackage{multirow}

\newcommand{\RN}[1]{\MakeUppercase{\romannumeral #1}}
\newcommand{\tool}{TRAVEL\xspace}

\newcommand{\modA}{Rule-Guided Reasoning Path Search\xspace}
\newcommand{\modB}{Dual-Reward RL Optimization\xspace}

\begin{document}

\title{Towards Reliable C-to-Rust Translation with Rule-Guided Reasoning and Reinforcement Learning}

\author{Feng Luo}
\affiliation{%
  \institution{Harbin Institute of Technology, Shenzhen}
  \country{China}
}
\email{hitszluofeng@foxmail.com}

\author{Jiachen Liu}
\affiliation{%
  \institution{Harbin Institute of Technology, Shenzhen}
  \country{China}
}
\email{jiachenliu@stu.hit.edu.cn}

\author{Cuiyun Gao}
\authornote{Corresponding author.}
\affiliation{%
  \institution{Harbin Institute of Technology, Shenzhen}
  \country{China}
}
\email{gaocuiyun@hit.edu.cn}

\author{Jia Feng}
\affiliation{%
  \institution{Harbin Institute of Technology, Shenzhen}
  \country{China}
}
\email{jiafeng@stu.hit.edu.cn}

\author{Kui Liu}
\affiliation{%
  \institution{Huawei Software Engineering Application Technology Lab}
  \country{China}
}
\email{kui.liu@huawei.com}

\renewcommand{\shortauthors}{Luo et al.}

\begin{abstract}
The migration of legacy C programs to Rust has become an important direction for improving software memory safety while alleviating the high cost of manual rewriting. 
Leveraging large language models (LLMs) for automated C-to-Rust translation has emerged as a promising direction. 
However, existing LLM-based approaches remain limited.
On the one hand, LLMs exhibit limited capability in identifying
Rust-specific rules, and inadequate handling of Rust syntax often results in incorrect translations.
On the other hand, existing LLMs often struggle to accurately capture the semantics of complex code, resulting in incorrect translations.

To address these challenges, we propose a \textbf{T}ranslation f\textbf{RA}mewo-rk \textbf{V}ia rule-guided reasoning and r\textbf{E}inforcement \textbf{L}earning, namely \tool, consisting of two modules.
The first module employs Monte Carlo Tree Search (MCTS)-based reasoning path construction guided by Rust-specific rules, steering the search toward translation steps that respect the syntactic rules that LLMs frequently violate.
The second module introduces reinforcement learning that couples execution feedback with reasoning-quality signals, encouraging the model to construct reasoning paths that accurately capture program semantics, thereby ensuring that the generated Rust code preserves the intended behavior of the original C program.

We evaluate \tool on three datasets: xCodeEval (a public benchmark), OS-Bench (functions collected from the Linux kernel), and HW-Bench (an industrial dataset from Huawei).
On xCodeEval, \tool outperforms all baselines across three backbone LLMs. 
In particular, compared to the strongest prompting baseline IRENE, \tool improves computational accuracy (CA) by 26.22\% and compilation success rate (CSR) by 18.77\%.
On HW-Bench and OS-Bench, \tool further improves CSR by 18.28\% and 16.51\%, respectively, while reducing unsafe rate (UR) by 13.06\% and 13.08\%, respectively.
\end{abstract}



\maketitle

\section{Introduction}
\label{sec:introduction}

Memory-safety vulnerabilities remain the dominant source of critical software flaws in modern systems, and are widely exploited across browsers, operating systems, and cloud infrastructure~\cite{cisa_memory_safety_2023, project_zero_memory_safety}.
Prior studies show that around 70\% of vulnerabilities in large codebases such as Microsoft products and Android stem from memory-safety issues in C and C++~\cite{miller2019trends, googleandroid2022memorysafe}.
Rust offers memory safety without garbage collection, while retaining the low-level control of C/C++ through its ownership and borrowing model~\cite{rustforlinux_nova}.
It has seen increasing adoption in practice, including in the Linux kernel, AWS Firecracker, and security-critical components of Windows~\cite{agache2020firecracker, microsoft2023rustwindows}.
However, manually migrating existing C projects to Rust is costly, as developers must learn Rust’s safety model and preserve the semantics of legacy code.
Therefore, automated C-to-Rust translation, which aims to generate semantically equivalent Rust code, has emerged as a promising direction.

Early work on C-to-Rust translation mainly relies on rule-based techniques~\cite{DBLP:conf/cav/ZhangDYW23,DBLP:conf/kbse/HongR24,DBLP:conf/icse/LingYWWCH22}, with C2Rust~\cite{immunant2022c2rust} as a representative system built on Clang~\cite{clang} and LLVM~\cite{DBLP:conf/cgo/LattnerA04}, and follow-up work extending it to specific scenarios such as pointer handling and parallel API migration~\cite{concrat,ownership,Emre2021,Emre2023}.
While efficient and interpretable, these methods depend heavily on handcrafted expert rules, making them labor-intensive to develop and limited in coverage when faced with the diverse patterns of real-world programs~\cite{Emre2021,Hong_static}.

Recently, large language models (LLMs) have shown strong potential in code generation and translation~\cite{DBLP:journals/corr/abs-2409-10506,DBLP:journals/corr/abs-2412-14234,DBLP:journals/corr/abs-2501-14257}, motivating studies that formulate C-to-Rust translation as a sequence-to-sequence task and learn patterns from large-scale code corpora~\cite{VERT,crownllm}. 
However, existing LLM-based methods still face two key challenges. 
First, LLMs often fail to follow Rust-specific rules, such as ownership, borrowing, and lifetimes (see Fig.~\ref{case1}). 
These rules are central to Rust's safety model, but they are not explicitly optimized during pretraining. 
As a result, LLMs may generate Rust code that violates compilation rules or relies on unsafe constructs to bypass safety checks. 
Existing methods that use external tools or static feedback~\cite{C2S,Multi} mainly repair errors after generation, and thus provide limited guidance during the translation process.
Second, LLMs often struggle to accurately capture the semantics of complex code (as shown in Fig.~\ref{case2}). 
C-to-Rust translation often depends on capturing subtle semantics in the source program. 
However, these semantics are not explicitly modeled during generation, making it difficult for LLMs to consistently reflect them in the generated code.

To address these challenges, we propose \tool, a translation framework via rule-guided reasoning and reinforcement learning, consisting of two key modules:

\begin{itemize}
    \item \textbf{\modA Module}: 
    This module constructs translation reasoning paths using MCTS guided by Rust-specific rules. 
    These rules capture constraints that LLMs often violate, such as ownership, borrowing and lifetimes. 
    By steering the search toward rule-compliant translation steps, this module produces higher-quality reasoning trajectories for subsequent training.

    \item \textbf{\modB Module}: 
    This module trains a reward model to evaluate the quality of translation reasoning paths, and then optimizes the translation policy with GRPO using both reasoning-quality rewards and execution feedback. 
    This encourages the model to preserve the semantics of the original C program while producing Rust code that follows Rust-specific rules.
\end{itemize}

To validate the performance of \tool, We conduct experiments on three datasets: xCodeEval~\cite{xcode}, a public function-level benchmark; HW-Bench, an industrial dataset from Huawei; and OS-Bench, a set of 107 C functions collected from the Linux kernel. 
The three datasets cover complementary domains: xCodeEval focuses on advanced programming and mathematics tasks, while HW-Bench and OS-Bench represent industrial and real-world code scenarios. 
On xCodeEval, \tool consistently outperforms all baselines across three backbone LLMs, achieving average gains of 35.61\% in computational accuracy (CA) and 30.08\% in compilation success rate (CSR). 
On HW-Bench and OS-Bench, \tool further improves CSR by 18.28\% and 16.51\%, while reducing unsafe rate (UR) by 13.06\% and 13.08\%, respectively. 
These results show that \tool is effective for both benchmark programs and real-world C-to-Rust migration scenarios.

In summary, our main contributions are as follows:
\begin{itemize}
    \item We identify two key challenges in LLM-based C-to-Rust translation: the limited ability of LLMs to follow Rust-specific rules, and the lack of mechanisms to improve reasoning quality for semantic preservation.
    
    \item We propose \tool, a novel LLM-based framework that combines rule-guided MCTS and reinforcement learning. 
    The \modA module discovers translation paths under Rust-specific rule guidance, while the \modB module improves semantic faithfulness using reasoning-quality rewards and execution feedback.
    
    \item We conduct extensive experiments on three datasets that cover a broad spectrum of domains, ranging from advanced programming and mathematical tasks to industrial applications and real-world system code.
    The results show that \tool consistently outperforms all competitive baselines under comparable model sizes. 
    Notably, our 7B model achieves performance competitive with strong proprietary models such as GPT-4o.
\end{itemize}


\begin{figure*}
    \centering
    \includegraphics[width=0.95\textwidth]{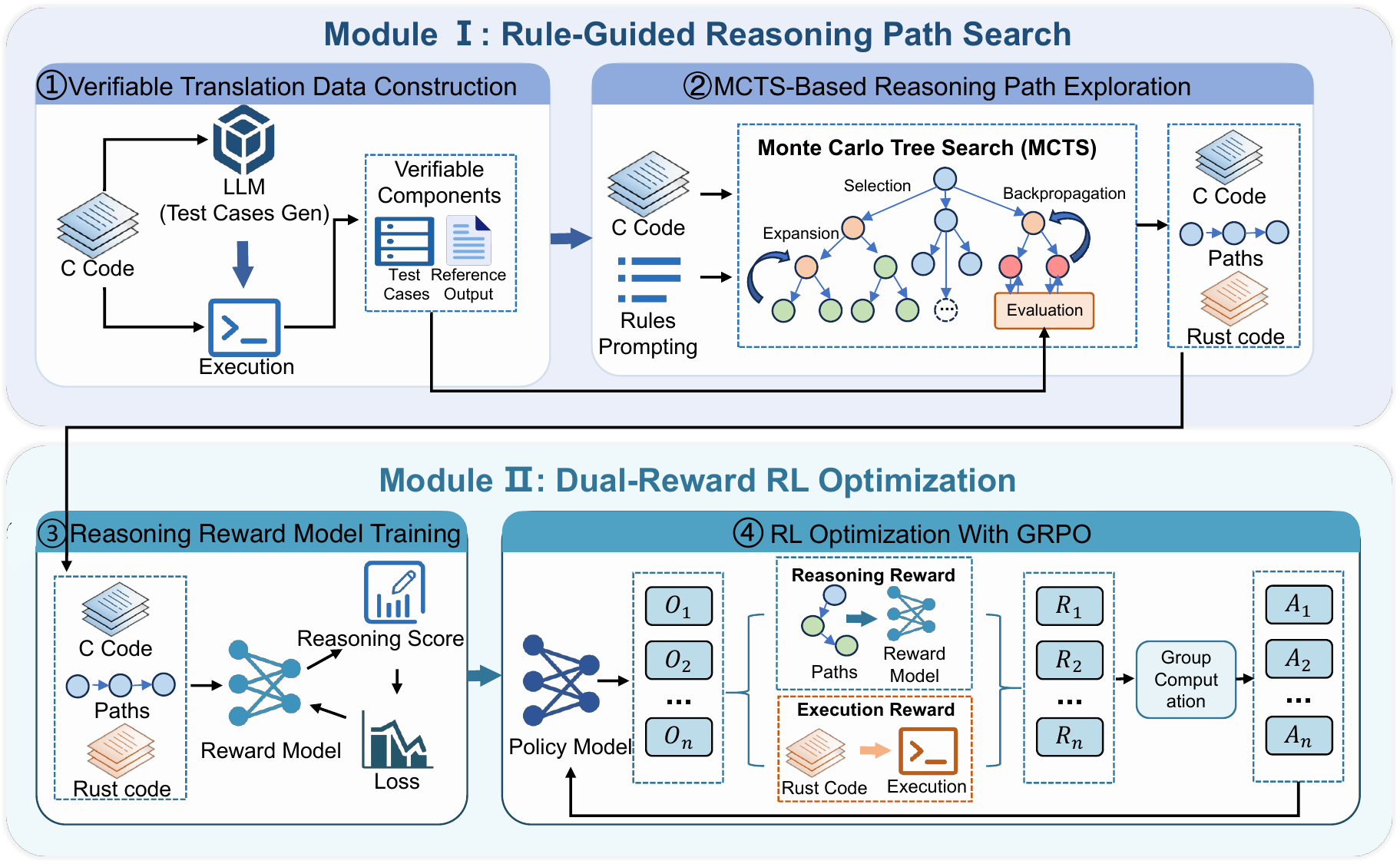}
    \caption{Overview of \tool. }
    \label{architecture}
\end{figure*}

\section{Methodology}
\label{sec:methodology}

In this section, we propose \tool, a translation framework via rule-guided reasoning path search and reinforcement learning. We first present the overview of \tool and then elaborate on its module details in the following subsections

\subsection{Overview}
\label{subsec:overview}

As illustrated in Fig.~\ref{architecture}, \tool consists of two main modules.
\modA Module applies MCTS with Rust rule-guided constraints to explore candidate reasoning paths and validates each path against our Verifiable Components (\S\ref{subsec:module_a}), propagating execution feedback through the search tree to collect high-quality training triplets $(C, P, R)$, where $C$ is the C code, $P$ is the reasoning path, and $R$ is the Rust code.
\modB Module uses these $(C, P, R)$ triplets to train a reward model that scores reasoning path quality, initializes the policy via supervised fine-tuning, and then optimizes it with GRPO using a combined reward of an \emph{execution reward} $r_{\text{exec}}$and a \emph{reasoning reward} $r_{\text{reason}}$, encouraging the model to produce translations that are both semantically correct and Rust-compliant.

\subsection{Module \RN{1}: \modA}
\label{subsec:module_a}

This module consists of two components. \textit{Verifiable Translation Data Construction} provides reliable feedback for semantic preservation and constructs the verifiable data. \textit{MCTS-Based Reasoning Path Exploration} enables the model to explicitly explore and compare alternative translation trajectories under rule-guided constraints.

\subsubsection{Verifiable Components Construction}

Evaluating whether a translated Rust program preserves the semantics of the original C code is essential but challenging, as most C-to-Rust datasets lack executable test suites. To address this, we introduce \textit{Verifiable Components}, which attach test inputs and reference outputs to each C function for semantic verification.

Given a C function, we prompt a strong LLM (Qwen2.5-Coder-32B-Instruct) to generate 5 test inputs, execute them on the original program, and collect the outputs. These input-output pairs form the Verifiable Components. To verify a Rust translation, we execute it on the same inputs and compare outputs with the references; the translation is considered correct if all outputs match.

Verifiable Components are used throughout \tool: Module~\RN{1} uses them to evaluate candidate reasoning paths during MCTS search, while Module~\RN{2} uses them to compute execution rewards for reinforcement learning, enabling semantic verification without pre-existing test cases.

\subsubsection{MCTS-Based Reasoning Path Exploration}

Monte Carlo Tree Search (MCTS) is a heuristic search algorithm that combines tree search with Monte Carlo sampling. It has gained prominence due to its success in applications such as AlphaGo and is particularly effective for sequential decision-making problems with large and complex state spaces where traditional methods become intractable~\cite{coulom2006efficient,silver2016mastering,silver2017mastering}. We adopt MCTS to explore diverse candidate reasoning paths for C-to-Rust translation, where each path corresponds to a sequence of reasoning steps that progressively transform a C function into its Rust counterpart.

Specifically, we formulate reasoning path generation as an MCTS search process. Each node represents an intermediate reasoning state consisting of the accumulated natural-language reasoning trace generated so far, while each action corresponds to generating the next reasoning step conditioned on the current reasoning state.

During MCTS, \tool starts with the initial state as the root and systematically grows the search tree by adding new nodes. Within the context of C-to-Rust translation, we customize the four key operations of the MCTS algorithm as follows:

\begin{table*}[t]
\caption{
    Selected Rust translation rules derived from the translation prompt used in \tool.
    Each rule is illustrated with a representative C-to-Rust transformation example. The complete rules can be found in our repository~\cite{ourrepo}.
}
\label{tab:taxonomy}
\resizebox{\textwidth}{!}{%
\begin{tabular}{l|p{4.8cm}|p{5.5cm}|p{5.5cm}}
\toprule
\textbf{Rule} & \textbf{Description} & \textbf{C Source} & \textbf{Rust Target} \\
\midrule

Replace malloc with Ownership
  & Use ownership types such as \texttt{Box<T>} or \texttt{Vec<T>} instead of \texttt{malloc}.
  & \texttt{int *p = malloc(sizeof(int)); *p = 42;}
  & \texttt{let mut p = Box::new(42i32);} \\
\midrule

Automatic Memory Management
  & Do not manually free memory; rely on Rust's ownership and drop system.
  & \texttt{free(p);}
  & \textit{(implicit drop when \texttt{p} goes out of scope)} \\
\midrule

NULL Return Handling
  & Convert functions returning \texttt{NULL} into \texttt{Option<T>} or \texttt{Result<T, E>}.
  & \texttt{int* find() \{ return NULL; \}}
  & \texttt{fn find() -> Option<i32> \{ None \}} \\
\midrule

Pointer to Reference
  & Prefer \texttt{\&T} / \texttt{\&mut T} over raw pointers when borrowing semantics allow.
  & \texttt{int x; int *p = \&x;}
  & \texttt{let x: i32 = 0; let p = \&x;} \\


\bottomrule
\end{tabular}%
}
\end{table*}

\textbf{Selection.} Starting from the root node $s_{0}$, the tree $\mathcal{T}_k$ is traversed by repeatedly selecting the child with the highest PUCT score~\cite{PUCT} until an unexpanded node is reached. PUCT uses the LLM's prior probability of each action together with visit-based statistics to balance exploitation of high-value paths and exploration of under-visited ones. Formally, at state $s_t$ we select
\begin{equation}
a_t = \arg\max_{a \in \mathcal{T}_k} \left[ \hat{Q}(s_t, a) + c_{\text{puct}} \, \pi_{\theta_k}(a \mid s_t)\sqrt{\frac{N_{\text{parent}}(a)}{1 + N(s_t, a)}} \right],
\label{eq:puct}
\end{equation}
where $\hat{Q}(s_t, a)$ is the estimated action value, $\pi_{\theta_k}(a \mid s_t)$ is the LLM's prior over actions, $N(s_t, a)$ is the visit count of $a$ at $s_t$, $N_{\text{parent}}(a)$ is the visit count of its parent, and $c_{\text{puct}}$ controls the exploration weight. To bias selection toward idiomatic Rust translations, we guide the LLM's action prior implicitly through rule-augmented prompts and few-shot examples

\textbf{Expansion.} At an unexpanded leaf $s_t$, we prompt the LLM to sample $K$ candidate next-step actions $\{a_t^{(1)}, \ldots, a_t^{(K)}\}$, each spawning a child node $s_{t+1}^{(i)}$ that extends the current partial translation. To obtain promising steps, in-context exemplars instruct the LLM to wrap every reasoning step between \texttt{<step>} and \texttt{</step>}, and we truncate generation at the first closing tag. The exemplars are deliberately designed to elicit a two-stage reasoning process within each path: the model first analyzes the source C function's semantics and then determines how to realize it in Rust by leveraging a Rust-specific translation rule library. Building upon prior C-to-Rust translation studies~\cite{Emre2021}, we summarize and extend this rule library (Table~\ref{tab:taxonomy}), which provides structured guidance for generating Rust-specific reasoning steps during path construction. This semantics-then-rules organization ensures that every candidate path jointly attends to behavioral correctness and Rust-specific idiomatic compliance, providing the dual-objective supervision later consumed by Module~\RN{2}.

\textbf{Evaluation.} When a node $s_t$ contains a complete Rust translation, we evaluate it with the source function's \textit{Verifiable Components} by executing the translation on the recorded test inputs and assigning a binary reward:
\begin{equation}
\hat{V}(s_t) = r =
\begin{cases}
+1, & \text{if all test cases pass}, \\
-1, & \text{if any test case fails},
\end{cases}
\label{eq:reward}
\end{equation}
where the test cases are those stored in the \textit{Verifiable Components} of the original C function.

\textbf{Backpropagation.} The reward $\hat{V}(s_t)$ obtained at a terminal node is propagated backward along the selected path, updating the visit counts and action value estimates of all ancestor nodes. Concretely, for each state-action pair $(s, a)$ on the path, the visit count is incremented as $N(s,a) \leftarrow N(s,a) + 1$, and the action value is updated by averaging the propagated rewards:
\begin{equation}
\hat{Q}(s,a) \leftarrow \frac{1}{N(s,a)} \sum_{i=1}^{N(s,a)} \hat{V}(s_t^{(i)}),
\end{equation}
where $\hat{V}(s_t^{(i)}) \in {+1, -1}$ is the terminal reward defined in Eq.~\eqref{eq:reward} for the $i$-th simulation passing through $(s,a)$. This process encourages future selections to favor trajectories that consistently lead to correct translations. Finally, we aggregate the mean of node value estimates along each reasoning path as its overall score, which is used to assess the quality of the generated reasoning trajectory.

After the search terminates, each explored path yields a training triplet $(C, P, R)$, where $C$ is the input C code, $P$ is the reasoning path with its estimated value, and $R$ is the resulting Rust translation.
These triplets serve as supervision data for the reinforcement learning stage in Module~\RN{2}.

\subsection{Module \RN{2}: \modB}
\label{subsec:module_b}

The goal of this module is to optimize the translation model using the search data collected in Module \RN{1}.
While the \modA module identifies promising reasoning paths for individual training instances, this module enables the model to internalize and reinforce such reasoning patterns for C-to-Rust translation.
To this end, we introduce a dual-reward reinforcement learning framework that jointly optimizes \emph{reasoning quality} and \emph{execution outcomes}.

Let the data collected from Module \RN{1} be represented as
\begin{equation}
\mathcal{D}=\{(C_i, P_i, R_i, v_i)\}_{i=1}^{N}
\end{equation}
where \(C_i\) denotes the input C code, \(P_i\) denotes the reasoning path explored by MCTS, \(R_i\) denotes the corresponding Rust translation generated along that path, and \(v_i\) denotes the reasoning path value estimated by MCTS through verification-based evaluation and backpropagation. 
Intuitively, \(v_i\) quantifies the quality of a reasoning path in terms of its ability to produce translations that satisfy the executable test cases.

\subsubsection{Reward Model Training}

We train a reward model to estimate the quality of reasoning paths and their resulting translations. 
Instead of explicit regression, we formulate reward prediction as a conditional generation task and train the model via supervised fine-tuning (SFT). 
Given a triplet $(C_i, P_i, R_i)$, the model is trained to generate the corresponding path value $v_i$ from Module~\RN{1}.

The objective follows the standard language modeling loss:
\begin{equation}
\mathcal{L}_{\text{rm}} 
= 
- \frac{1}{N} \sum_{i=1}^{N} \log p_{\phi}(v_i \mid C_i, P_i, R_i)
\end{equation}
This formulation can be viewed as \emph{implicit regression via generative modeling}, where continuous reward signals are learned through next-token prediction~\cite{mahan2025generative, akhauri2025performance}. 
Compared to execution-based supervision alone, it provides a denser training signal for evaluating reasoning quality despite sparse binary outcomes from compilation and testing.

\subsubsection{Policy Initialization}

Before reinforcement learning, we obtain an initial policy model \(\pi_{\theta}\) through a cold-start training stage. 
Specifically, the model is first supervised on the high-quality search data collected from Module \RN{1}, so that it learns to generate Rust code conditioned on structured reasoning paths. 
Formally, for each training example \((C_i, P_i, R_i)\), the policy is optimized with the standard autoregressive objective:
\begin{equation}
\mathcal{L}_{\text{sft}}
=
-\frac{1}{N}
\sum_{i=1}^{N}
\log \pi_{\theta}(P_i, R_i \mid C_i).
\end{equation}
This initialization provides a reasonable behavioral prior over the task, enabling well-formed reasoning paths and translations.

\subsubsection{Dual-Reward GRPO Optimization}

After cold-start initialization, we further optimize the policy model using Group Relative Policy Optimization (GRPO)~\cite{guo2025deepseek}. 

\paragraph{Optimization Objective.}
For each input C function $C$, GRPO~\cite{guo2025deepseek} samples a group of $G$ outputs $\{o^{(k)}\}_{k=1}^{G}$ from the previous policy $\pi_{\theta_{\mathrm{old}}}$, where each $o^{(k)} = (P^{(k)}, R^{(k)})$ pairs a reasoning path with its resulting Rust translation. The current policy $\pi_{\theta}$ is then updated by maximizing
\begin{equation}
\begin{aligned}
\mathcal{J}_{\text{GRPO}}(\theta)
= \,& \mathbb{E}_{C,\, \{o^{(k)}\} \sim \pi_{\theta_{\mathrm{old}}}(\cdot \mid C)} \\
& \frac{1}{G} \sum_{k=1}^{G} \Big(
\min\!\big(
\rho_k(\theta)\, A^{(k)},\;
\mathrm{clip}(\rho_k(\theta),\, 1-\varepsilon,\, 1+\varepsilon)\, A^{(k)}
\big) \\
& - \beta\, \mathbb{D}_{\mathrm{KL}}\!\left(\pi_{\theta} \,\|\, \pi_{\theta_{\mathrm{ref}}}\right)
\Big),
\end{aligned}
\label{eq:grpo_objective}
\end{equation}
where $\rho_k(\theta) = \pi_{\theta}(o^{(k)} \mid C) / \pi_{\theta_{\mathrm{old}}}(o^{(k)} \mid C)$ is the importance ratio, $\varepsilon$ is the clipping range, and $\beta$ controls the KL penalty strength.

The advantage $A^{(k)}$ is computed by normalizing rewards within the sampled group, eliminating the need for a separate value network:
\begin{equation}
A^{(k)} = \frac{r^{(k)} - \mathrm{mean}(\{r^{(j)}\}_{j=1}^{G})}{\mathrm{std}(\{r^{(j)}\}_{j=1}^{G})},
\label{eq:advantage}
\end{equation}
where $r^{(k)} = r(C, P^{(k)}, R^{(k)})$ is the combined reward defined above.

The KL divergence is estimated via the unbiased estimator:
\begin{equation}
\mathbb{D}_{\mathrm{KL}}\!\left(\pi_{\theta} \,\|\, \pi_{\theta_{\mathrm{ref}}}\right)
= \frac{\pi_{\theta_{\mathrm{ref}}}(o^{(k)} \mid C)}{\pi_{\theta}(o^{(k)} \mid C)}
- \log \frac{\pi_{\theta_{\mathrm{ref}}}(o^{(k)} \mid C)}{\pi_{\theta}(o^{(k)} \mid C)} - 1,
\label{eq:kl_div}
\end{equation}
where $\pi_{\theta_{\mathrm{ref}}}$ is the SFT-tuned reference policy obtained before reinforcement learning. This term keeps the updated policy from drifting excessively from the reference, ensuring training stability.

\paragraph{Execution Reward.}
The execution reward evaluates the final Rust translation based on executable outcomes. 
Given a generated Rust program \(R\), we first check whether it compiles successfully. 
If compilation fails, a penalty of \(-1\) is assigned. 
Otherwise, we execute \(R\) on the test cases derived from the \textit{Verifiable Components} of the original C program, and compute the pass rate:
\begin{equation}
\mathrm{PassRate}(R) = \frac{n_{\text{pass}}}{n_{\text{total}}},
\end{equation}

where \(n_{\text{pass}}\) and \(n_{\text{total}}\) denote the number of passed and total test cases, respectively. 
If no test case is passed, we assign a reward of \(-1\); otherwise, the reward equals the pass rate. 
Formally,
\begin{equation}
r_{\text{exec}}(R)
=
\begin{cases}
-1, & \text{if } R \text{ fails to compile}, \\
-1, & \text{if } \mathrm{PassRate}(R) = 0, \\
\mathrm{PassRate}(R), & \text{otherwise}.
\end{cases}
\end{equation}
This design enforces strong penalties for syntactically invalid or completely incorrect programs, while providing graded feedback for partially correct translations.

\paragraph{Reasoning Reward.}
The reasoning reward evaluates the quality of the generated reasoning path and its associated translation using the learned reward model:
\begin{equation}
r_{\text{reason}}(C, P, R)
=
f_{\phi}(C, P, R).
\end{equation}
Unlike the execution reward, which only reflects the quality of the final output, the reasoning reward provides additional supervision on the intermediate translation process. 
It encourages the model to generate structured and rules-aware reasoning paths that are more likely to lead to reliable translations.

\paragraph{Final Reward.}
We combine the two reward components into a unified reward:
\begin{equation}
r(C, P, R)
=
\lambda_{\text{exec}} \, r_{\text{exec}}(R)
+
\lambda_{\text{reason}} \, r_{\text{reason}}(C, P, R),
\end{equation}
where \(\lambda_{\text{exec}}\) and \(\lambda_{\text{reason}}\) are balancing coefficients. 
This design allows the policy to optimize both the final translation quality and the reasoning process that leads to it.

Through this dual-reward GRPO optimization, the policy model is encouraged not only to produce correct final translations, but also to generate reasoning paths that are more likely to lead to semantically faithful and Rust-compliant Rust programs. 
In this way, Module~\RN{2} complements Module~\RN{1}: guided by carefully designed few-shot exemplars, the reasoning paths in Module~\RN{1} first reason about the source program's semantics and then apply Rust-specific rules, so that each path jointly strengthens semantic understanding and syntactic handling; Module~\RN{2}'s reinforcement learning further consolidates both capabilities in the translation model itself.


\section{Experimental Setup}
\label{sec:experimental_setup}

\subsection{Research Questions}
\label{subsec:rqs}

To evaluate the effectiveness of \tool, we investigate the following research questions:

\begin{description}
    \item[\textbf{RQ1:}] How effective is \tool across different model scales for C-to-Rust translation?
    \item[\textbf{RQ2:}] How do the major components of \tool contribute to the final performance?
    \item[\textbf{RQ3:}] How does \tool perform in industrial scenarios?
\end{description}

\subsection{Datasets}
\label{subsec:datasets}
\textbf{Training Set.}
We construct a training dataset based on the training split of xCodeEval~\cite{xcode}, comprising 16,639 samples. Each sample consists of the original C source code, a sequence of reasoning paths produced by the \modA module of \tool in Section~\ref{sec:methodology}, and the corresponding Rust translation, where both the C and Rust programs have been validated against test cases. We partition the data into training and validation sets at a 9:1 ratio. 
\begin{table}[t]
\centering
\small
\setlength{\tabcolsep}{3pt}
\caption{Code characteristics statistics of HW-Bench and OS-Bench.}
\label{tab:stat}
\begin{tabular}{l|ccccc}
\toprule
Benchmark 
& \#Samp. 
& Avg. LOC 
& Avg. Depth 
& Max Depth 
& Avg. Derefs. \\
\midrule
HW-Bench & 100 & 15.33 & 0.611 & 0.890 & 2.22 \\
OS-Bench & 107 & 26.02 & 0.607 & 0.851 & 4.45 \\
\bottomrule
\end{tabular}
\end{table}

\textbf{Evaluation Set.}
We construct three evaluation sets to comprehensively assess the effectiveness of \tool. 
First, we select 515 samples from the evaluation split of xCodeEval, where each C code sample passes all test cases and has a corresponding Rust implementation. To further evaluate the applicability of \tool in real-world open-source and industrial scenarios, we build two additional benchmarks. The first one, OS-Bench, contains 107 open-source C functions collected from the Linux kernel repository~\cite{github_os_data}. We choose Linux kernel code because it represents large-scale, production-quality system software with complex control flow, low-level memory manipulation, and strict requirements on correctness and safety. These characteristics make it highly aligned with practical enterprise-level C-to-Rust migration demands, while also posing substantial challenges for automated translation. The second one, HW-Bench, consists of 100 C functions randomly sampled from proprietary projects provided by Huawei, covering diverse industrial scenarios. 
Table~\ref{tab:stat} further reports the statistics of code characteristics for both benchmarks.

\subsection{Baseline Methods}
\label{subsec:baselines}

In C-to-Rust translation experiments, we compare \tool with baselines as detailed below. \textbf{Instruction} provides the C code along with the prompt ‘Translate the following C code to Rust’ without any additional demonstration. \textbf{In-context Learning (ICL)}~\cite{ICL} includes several translation examples in the prompt to help the model generate the target Rust code. \textbf{Retrieval-augment generation (RAG)}~\cite{RAG} retrieves examples from the corpus similar to the target translation C code and includes them in the input prompt. \textbf{Vert}~\cite{VERT} uses an LLM with a two-stage refinement process. It first applies compiler-suggested fixes to the source code, then feeds the updated code and diagnostics back into the LLM for further correction. \textbf{IRENE}~\cite{IRENE} integrates Rust-specific rules and source-code semantics into an LLM-based C-to-Rust translation pipeline, using rule-guided prompting to steer the model toward safer and more idiomatic translations. \textbf{SFT}~\cite{SFT} fine-tunes the backbone model on our training set using standard supervised learning. 
It can be viewed as an ablated variant of \tool that removes the reinforcement learning stage.

\subsection{Implementation Details}
\label{subsec:implementation}


\textbf{Model selection and setup.}
We use Qwen2.5-Coder-Instruct at 7B and 3B scales~\cite{hui2024qwen2} and CodeGemma-7B-Instruct~\cite{codegemma} as the backbone models for training.
To provide a comprehensive comparison, we further include larger open-source models, including Qwen2.5-Coder-14B/32B-Instruct and Qwen3-Coder-30B-Instruct~\cite{yang2025qwen3}, as well as closed-source proprietary models such as GPT-4o, GPT-4o-mini~\cite{openai2024gpt4technicalreport}, GPT-5.4-mini~\cite{openai_gpt5}, and Claude-Haiku-4.5~\cite{2023claude}.
We download open-source models from HuggingFace~\cite{wolf2020transformers} and utilize PyTorch~\cite{paszke2019pytorch} for local deployment and inference, while closed-source models are accessed via their official APIs.

\textbf{Implementation of baselines.}
For ICL, we follow the prior work~\cite{DBLP:conf/kbse/GaoWGWZL23} and randomly sample four examples from the corpus.
For RAG, we retrieve the most relevant example using the BM25 algorithm.
For Vert, we directly use the replication packages released by the authors and adapt them to our tasks.
For IRENE, we select the example with the highest similarity and apply one iteration of refinement. 
For \tool, the final reward is computed with $\lambda_{exec} = \lambda_{reason} = 0.5$, and Qwen2.5-Coder-32B-Instruct is used as the backbone model for MCTS.

\textbf{Hyperparameter and environment settings.}
For the MCTS-based reasoning path exploration module, we set the maximum number of reasoning steps per search episode to 5.
For the GRPO-based reinforcement learning training, we use the VERL framework~\cite{sheng2024hybridflow} with the following hyperparameters:
the number of candidate samples per prompt $G = 8$, 
the KL divergence coefficient $\beta = 0.001$, 
the clipping range $\varepsilon = 0.2$, 
the learning rate = $1 \times 10^{-6}$, 
a training batch size of 32 prompts (each with 8 sampled responses), 
2 training epochs, 
and a maximum sequence length of 4096 tokens.
We adopt LoRA~\cite{hu2022lora} for parameter-efficient training with rank $r = 32$, scaling factor $\alpha = 64$, and dropout rate = 0.
As for the hyperparameters of generation, we configure all LLMs with greedy decoding (temperature = 0, top-p = 1) to produce deterministic outputs.
We execute Python programs with Python 3.10.0 and compile all Rust programs with Rust 1.81.0.
All experiments are conducted on an Ubuntu-20.04 server equipped with four NVIDIA A100 GPUs.

\subsection{Performance Metrics}
\label{subsec:metrics}

To evaluate the translation accuracy, we use the following two widely used performance metrics~\cite{yang2024} in our evaluation:

\textbf{Computational Accuracy (CA)} 
evaluates whether the candidate translation generates the same outputs as the reference when given the same inputs. It can be formally defined as:
\begin{align}
\text{CA} &= \frac{\sum_{k=1}^{N} ca(y_k, \hat{y}_k)}{N} \\
ca(y_k, \hat{y}_k) &=
\begin{cases}
1, & \text{if } Exec_k(y_k) = Exec_k(\hat{y}_k) \\
0, & \text{otherwise}
\end{cases}
\end{align}
where $N$ is the total number of evaluated code samples. $y_k$ and $\hat{y}_k$ represent the ground truth and the generated translation for the $k$-th sample, respectively. $E_{x \in C_k}(\cdot)$ denotes the execution result of a program on the $k$-th test case. The indicator function $\mathrm{ca}(y_k, \hat{y}_k)$ equals 1 if both outputs match, and 0 otherwise.

\textbf{Compilation Success Rate (CSR)}
measures the percentage of translated programs that successfully compile without errors.

To evaluate the translation safety, following prior work~\cite{Multi}, we use the metrics below:

\textbf{Unsafe Rate (UR)} refers to the proportion of unsafe samples in the dataset. 

\textbf{Unsafe Loc Rate (ULR)} denotes the ratio of unsafe lines of code within each individual sample.


\section{Experimental Results}
\label{sec:experimental_result}

\subsection{RQ1: How effective is \tool across different model scales for C-to-Rust translation?}
\label{subsec:rq1}

To evaluate the overall effectiveness of \tool, we compare \tool with multiple baselines on xCodeEval benchmark. Table~\ref{tab:rq1-table} reports the results, from which we derive the following findings:

\begin{table*}[]
\caption{Comparison results with baseline models in terms of translation accuracy (CA, CSR) and translation safety (UR, ULR). The bold figures indicate the best results.}
\label{tab:rq1-table}
\small
\begin{tabular}{@{}ccccccc@{}}
\toprule
\multicolumn{2}{c|}{\multirow{2}{*}{Approach}} & \multicolumn{1}{c|}{\multirow{2}{*}{Size/Version}} & \multicolumn{2}{c|}{Translation Accuracy} & \multicolumn{2}{c}{Translation Safety} \\
\multicolumn{2}{c|}{} & \multicolumn{1}{c|}{} & CA & \multicolumn{1}{c|}{CSR} & UR & ULR \\ \midrule
\multicolumn{7}{c}{Closed-source Proprietary Models} \\ \midrule
\multicolumn{2}{c|}{GPT-5.4-mini} & \multicolumn{1}{c|}{Mar. 2026} & 96.31 & \multicolumn{1}{c|}{99.03} & 5.83 & 2.75 \\
\multicolumn{2}{c|}{GPT-4o} & \multicolumn{1}{c|}{Nov. 2024} & 69.51 & \multicolumn{1}{c|}{93.59} & 5.05 & 2.35 \\
\multicolumn{2}{c|}{GPT-4o-mini} & \multicolumn{1}{c|}{July. 2024} & 58.25 & \multicolumn{1}{c|}{85.63} & 5.05 & 1.73 \\
\multicolumn{2}{c|}{Claude-Haiku-4.5} & \multicolumn{1}{c|}{Oct. 2025} & 74.37 & \multicolumn{1}{c|}{95.34} & 3.11 & 1.04 \\ \midrule
\multicolumn{7}{c}{15B+ Models} \\ \midrule
\multicolumn{2}{c|}{Qwen2.5-Coder-32B-Instruct} & \multicolumn{1}{c|}{32B} & 59.03 & \multicolumn{1}{c|}{87.18} & 0.97 & 0.36 \\
\multicolumn{2}{c|}{Qwen2.5-Coder-14B-Instruct} & \multicolumn{1}{c|}{14B} & 45.44 & \multicolumn{1}{c|}{74.17} & 2.52 & 0.67 \\
\multicolumn{2}{c|}{Qwen3-Coder-30B-Instruct} & \multicolumn{1}{c|}{30B} & 52.04 & \multicolumn{1}{c|}{76.50} & 4.27 & 1.36 \\ \midrule
\multicolumn{7}{c}{7B-level And 3B-level Models} \\ \midrule
\multirow{7}{*}{Qwen2.5-Coder-7B-Instruct} & \multicolumn{1}{c|}{Instruction} & \multicolumn{1}{c|}{7B} & 38.83 & \multicolumn{1}{c|}{67.57} & 3.11 & 0.49 \\
~ & \multicolumn{1}{c|}{ICL} & \multicolumn{1}{c|}{7B} & 34.95 & \multicolumn{1}{c|}{63.50} & 0.78 & 0.10 \\
~ & \multicolumn{1}{c|}{RAG} & \multicolumn{1}{c|}{7B} & 34.56 & \multicolumn{1}{c|}{63.69} & 0.58 & 0.17 \\
~ & \multicolumn{1}{c|}{Vert} & \multicolumn{1}{c|}{7B} & 37.28 & \multicolumn{1}{c|}{76.70} & 3.11 & 0.86 \\
~ & \multicolumn{1}{c|}{IRENE} & \multicolumn{1}{c|}{7B} & 51.84 & \multicolumn{1}{c|}{83.88} & 1.36 & 0.12 \\
~ & \multicolumn{1}{c|}{SFT} & \multicolumn{1}{c|}{7B} & 68.35 & \multicolumn{1}{c|}{\textbf{87.77}} & \textbf{0.00} & \textbf{0.00} \\
~ & \multicolumn{1}{c|}{\tool} & \multicolumn{1}{c|}{7B} & \textbf{71.07} & \multicolumn{1}{c|}{87.38} & \textbf{0.00} & \textbf{0.00} \\ \midrule
\multirow{7}{*}{Codegemma-7B-Instruct} & \multicolumn{1}{c|}{Instruction} & \multicolumn{1}{c|}{7B} & 4.08 & \multicolumn{1}{c|}{10.87} & 3.50 & 0.32 \\
~ & \multicolumn{1}{c|}{ICL} & \multicolumn{1}{c|}{7B} & 19.42 & \multicolumn{1}{c|}{36.12} & \textbf{0.39} & \textbf{0.01} \\
~ & \multicolumn{1}{c|}{RAG} & \multicolumn{1}{c|}{7B} & 9.71 & \multicolumn{1}{c|}{24.85} & 0.58 & 0.05 \\
~ & \multicolumn{1}{c|}{Vert} & \multicolumn{1}{c|}{7B} & 2.33 & \multicolumn{1}{c|}{9.51} & 1.75 & 0.13 \\
~ & \multicolumn{1}{c|}{IRENE} & \multicolumn{1}{c|}{7B} & 32.43 & \multicolumn{1}{c|}{52.82} & 0.78 & 0.02 \\
~ & \multicolumn{1}{c|}{SFT} & \multicolumn{1}{c|}{7B} & 66.99 & \multicolumn{1}{c|}{86.21} & 0.78 & 0.03 \\
~ & \multicolumn{1}{c|}{\tool} & \multicolumn{1}{c|}{7B} & \textbf{70.68} & \multicolumn{1}{c|}{\textbf{86.99}} & 0.78 & 0.03 \\ \midrule
\multirow{7}{*}{Qwen2.5-Coder-3B-Instruct} & \multicolumn{1}{c|}{Instruction} & \multicolumn{1}{c|}{3B} & 18.06 & \multicolumn{1}{c|}{36.31} & 0.19 & 0.01 \\
~ & \multicolumn{1}{c|}{ICL} & \multicolumn{1}{c|}{3B} & 20.19 & \multicolumn{1}{c|}{47.38} & \textbf{0.00} & \textbf{0.00} \\
~ & \multicolumn{1}{c|}{RAG} & \multicolumn{1}{c|}{3B} & 24.66 & \multicolumn{1}{c|}{52.43} & \textbf{0.00} & \textbf{0.00} \\
~ & \multicolumn{1}{c|}{Vert} & \multicolumn{1}{c|}{3B} & 4.27 & \multicolumn{1}{c|}{22.14} & 6.99 & 0.62 \\
~ & \multicolumn{1}{c|}{IRENE} & \multicolumn{1}{c|}{3B} & 37.86 & \multicolumn{1}{c|}{56.50} & \textbf{0.00} & \textbf{0.00} \\
~ & \multicolumn{1}{c|}{SFT} & \multicolumn{1}{c|}{3B} & 57.86 & \multicolumn{1}{c|}{\textbf{77.48}} & \textbf{0.00} & \textbf{0.00} \\
~ & \multicolumn{1}{c|}{\tool} & \multicolumn{1}{c|}{3B} & \textbf{59.03} & \multicolumn{1}{c|}{75.15} & \textbf{0.00} & \textbf{0.00} \\ \bottomrule
\end{tabular}
\end{table*}

\textbf{Simple prompting yields limited gains in LLMs' translation ability for C-to-Rust.}
As shown in Table~\ref{tab:rq1-table}, the prompting-based baselines have limited effectiveness in enhancing model translation capabilities. For instance, Qwen2.5-Coder-7B-Instruct with IRENE only reaches 51.84\% CA and 83.88\% CSR, while the remaining prompting variants cluster between 34.56\% and 38.83\% CA. Moreover, prompting baselines generally perform poorly on Codegemma-7B-Instruct; with the exception of IRENE, all other methods achieve at most 19.42\% CA and 36.12\% CSR. These results suggest that, without parameter-level adaptation, neither in-context examples nor retrieval can teach the model to internalize Rust's semantic and syntactic constraints, the dual challenge that motivates \tool.

\textbf{\tool effectively improves models' translation capabilities in C-to-Rust, particularly in semantic understanding and Rust-specific rules syntactic handling.}
As shown in Table~\ref{tab:rq1-table}, models augmented with \tool demonstrate substantial improvements across all metrics. The model achieves strong performance after SFT, and further improvements are obtained through subsequent reinforcement learning.
For Codegemma-7B-Instruct, applying \tool improves performance by 48.19\% and 50.26\% in terms of CA and CSR on average across all baselines. Notably, \tool outperforms the state-of-the-art baseline SFT by 3.69\% and 0.78\% on the same metrics. Similarly, for Qwen2.5-Coder-7B-Instruct, \tool yields consistent improvements, achieving average gains of 26.77\% and 13.53\% in CA and CSR, respectively.
Both SFT and the \tool substantially outperform all prompting baselines, confirming the strength in jointly modeling source semantics and Rust-specific syntactic rules. 

We observe that on Qwen2.5-Coder-3B/7B-Instruct, \tool yields slightly lower CSR than SFT. 
This is because our final reward optimizes semantic correctness rather than CSR alone. 
Although compilation failures are penalized, the execution reward mainly provides graded feedback through test-pass outcomes, while the reasoning reward encourages more semantically faithful reasoning paths. 
This can make the policy less conservative than SFT, occasionally introducing compilation errors while improving behavioral correctness.

\textbf{Small models augmented with \tool outperform some larger models.}
By comparing models of different sizes, we find that smaller models enhanced with \tool can outperform substantially larger ones. For example, Qwen2.5-Coder-7B-Instruct+\tool achieves 71.07\% CA and 87.38\% CSR on xCodeEval, outperforming Qwen2.5-Coder-32B-Instruct (59.03\% CA), Qwen3-Coder-30B-Instruct (52.04\% CA), and the proprietary GPT-4o-mini (58.25\% CA), while closing most of the gap to GPT-4o (69.51\% CA). Similarly, Qwen2.5-Coder-3B-Instruct+\tool attains 59.03\% CA, on par with Qwen2.5-Coder-32B-Instruct that has roughly 10 times more parameters. These results highlight that through \tool, smaller models can achieve comparable or superior performance to much larger models in C-to-Rust translation.

\textbf{Translation safety.}
Beyond accuracy, \tool also significantly improves translation safety. For instance, on Qwen2.5-Coder-7B-Instruct, \tool reduces UR and ULR by 1.49\% and 0.29\% on average compared to all baselines. Notably, it also demonstrates substantial advantages over strong closed-source models; for example, on CodeGemma-7B-Instruct, \tool achieves average reductions of 3.98\% and 1.94\% in UR and ULR, respectively, compared to the four closed-source models shown in Table~\ref{tab:rq1-table}.
This indicates that the rule-guided search effectively internalizes Rust safety constraints during training.

\begin{tcolorbox}
\textbf{Answer to RQ1:} 
\tool consistently improves C-to-Rust translation across backbones of different families and scales: it outperforms all same-scale baselines, matches or exceeds open-source models up to an order of magnitude larger, and narrows the gap to leading proprietary LLMs, while producing markedly safer Rust code with near-zero unsafe usage.
\end{tcolorbox}

\subsection{RQ2: How do the major components of \tool contribute to the final performance?}
\label{subsec:rq2}

To quantify the contribution of the major components of \tool, we conduct an ablation study across all three backbone models, as shown in Table~\ref{tab:rq2-table}:

\begin{itemize}
    \item \textbf{-w/o Reasoning Reward:} remove the reasoning-path quality reward, retaining only the execution reward during RL.
    \item \textbf{-w/o Execution Reward:} remove the execution reward, retaining only the reasoning-path quality reward during RL.
    \item \textbf{-w/o RL (SFT with Paths):} disable the RL stage and perform supervised fine-tuning on the MCTS-constructed reasoning paths.
    \item \textbf{-w/o RL \& Paths (SFT w/o Paths):} disable both the RL stage and reasoning-path construction, and perform supervised fine-tuning only on input–output pairs.
\end{itemize}

\begin{table*}[]
\caption{Ablation study of the major components of \tool on three backbone models.}
\label{tab:rq2-table}
\small
\begin{tabular}{@{}l|cc|cc|cc@{}}
\toprule
\multicolumn{1}{c|}{\multirow{2}{*}{Approach}} & \multicolumn{2}{c|}{Qwen2.5-Coder-3B-Instruct} & \multicolumn{2}{c|}{Qwen2.5-Coder-7B-Instruct} & \multicolumn{2}{c}{Codegemma-7B-Instruct} \\
\multicolumn{1}{c|}{} & CA & CSR & CA & CSR & CA & CSR \\ \midrule
-w/o Reasoning Reward & 58.06 & \textbf{77.48} & 70.29 & \textbf{88.74} & 69.71 & \textbf{87.57} \\
-w/o Execution Reward & 55.92 & 71.65 & 70.10 & 88.54 & 68.35 & 86.99 \\
-w/o RL (SFT with Paths) & 57.86 & 77.48 & 68.35 & 87.77 & 66.99 & 86.21 \\
-w/o RL \& Paths (SFT w/o Paths) & 48.16 & 70.87 & 67.18 & 83.30 & 65.63 & 82.52 \\ \midrule
\tool & \textbf{59.03} & 75.15 & \textbf{71.07} & 87.38 & \textbf{70.68} & 86.99 \\ \bottomrule
\end{tabular}
\end{table*}

\begin{figure}[t]
    \centering
    \includegraphics[width=\linewidth]{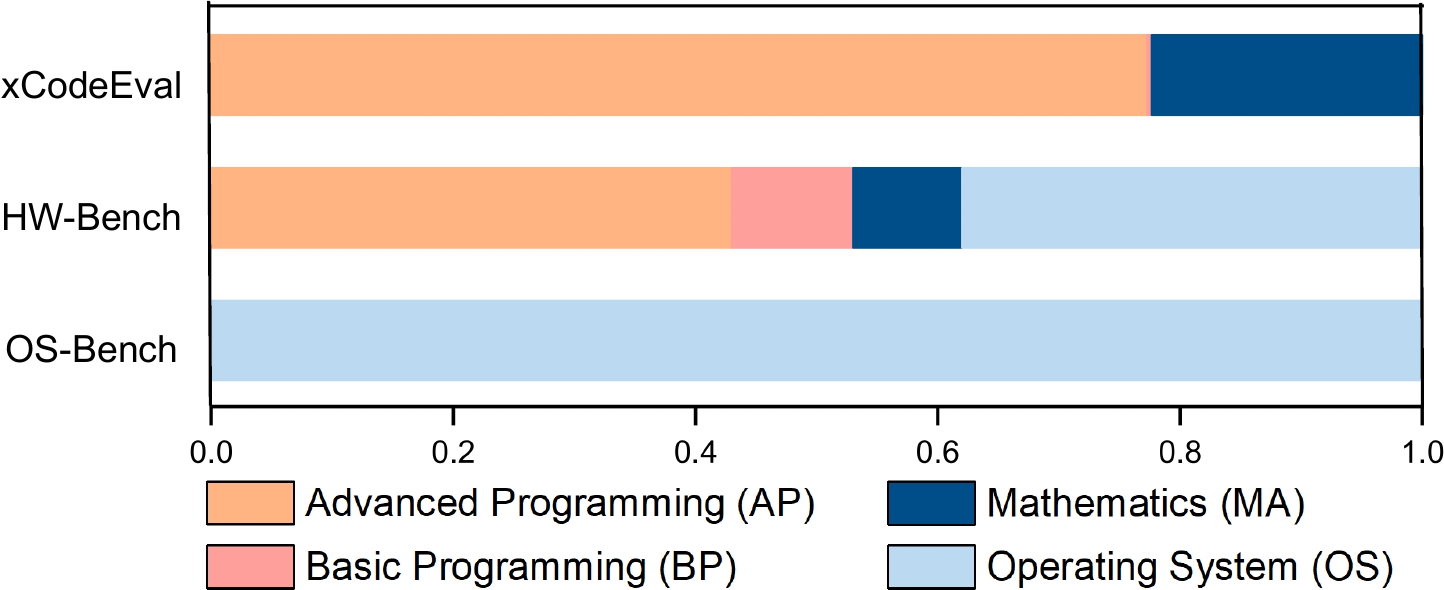}
    \caption{Application domain distributions of xCodeEval, HW-Bench and OS-Bench.}
    \label{domain}
\end{figure}

\textbf{Effect of the reasoning reward.}
As shown in Table~\ref{tab:rq2-table}, removing the reasoning reward (\textit{-w/o Reasoning Reward}) consistently lowers CA across all three backbones. For example, CA drops from 71.07\% to 70.29\% on Qwen2.5-Coder-7B-Instruct and from 70.68\% to 69.71\% on Codegemma-7B-Instruct, confirming that the reasoning paths encouraged by this reward are highly effective for capturing source semantics. Interestingly, CSR increases on all backbones in this setting. 
This is because removing the reasoning reward simplifies the optimization signal, leaving the model to rely primarily on execution-based feedback. 
As a result, the model tends to produce more conservative translations that prioritize compilability, often at the expense of semantic alignment.

\textbf{Effect of the execution reward.}
As shown in Table~\ref{tab:rq2-table}, removing the execution reward (\textit{-w/o Execution Reward}) leads to a consistent degradation in performance across all three backbones. For example, on Qwen2.5-Coder-3B-Instruct, CA drops from 59.03\% to 55.92\% and CSR falls from 75.15\% to 71.65\%. This confirms that execution-level feedback provides a concrete and verifiable signal that complements the more abstract reasoning-path reward and anchors the optimization in observable behavior.

\textbf{Effect of reinforcement learning.}
As shown in Table~\ref{tab:rq2-table}, removing the dual-reward RL stage (\textit{-w/o RL (SFT with Paths)}) results in performance declines across all three backbones. 
For example, CA drops from 71.07\% to 68.35\% on Qwen2.5-Coder-7B-Instruct and from 70.68\% to 66.99\% on Codegemma-7B-Instruct. 
The marginal CSR rise on the two Qwen backbones reflects a shift in optimization behavior: without RL, the model produces more conservative code that favors compilation, while dual-reward RL encourages semantically richer reasoning that may occasionally introduce compilation errors. 
Overall, this trade-off is favorable, as CA gains outweigh the small CSR fluctuations.

\textbf{Effect of reasoning-path construction.}
Comparing \textit{SFT w/o Paths} against \textit{SFT with Paths} isolates the contribution of the MCTS-constructed reasoning paths. Across all three backbones, removing the reasoning paths degrades both metrics. For example, on Qwen2.5-Coder-7B-Instruct, CA drops from 68.35\% to 67.18\% and CSR drops from 87.77\% to 83.30\%, indicating that the structured reasoning paths supply explicit semantic and rule-level guidance that plain input--output pairs cannot offer.

\begin{tcolorbox}
\textbf{Answer to RQ2:}
The major components of \tool jointly drive the final performance. Reasoning-path construction contributes the largest single gain, especially on smaller backbones, and the RL stage brings further consistent improvements in CA across all backbones. The reasoning-path and execution rewards play complementary roles: the former favors semantic validity, while the latter balances semantic validity and compilability, and combining them yields the best overall balance on larger backbones.
\end{tcolorbox}

\subsection{RQ3: How does \tool perform in industrial scenarios?}
\label{subsec:rq3}
\begin{table*}[]
\caption{Performance comparison on HW-Bench and OS-Bench}
\label{tab:rq3}
\small
\setlength{\tabcolsep}{4pt}
\begin{tabular}{@{}c|cccccc|cccccc|cccccc@{}}
\toprule
\multirow{3}{*}{Approach} & \multicolumn{6}{c|}{Qwen2.5-Coder-3B-Instruct} & \multicolumn{6}{c|}{Qwen2.5-Coder-7B-Instruct} & \multicolumn{6}{c}{Codegemma-7B-Instruct} \\ \cmidrule(l){2-19} 
 & \multicolumn{3}{c|}{HW-Bench} & \multicolumn{3}{c|}{OS-Bench} & \multicolumn{3}{c|}{HW-Bench} & \multicolumn{3}{c|}{OS-Bench} & \multicolumn{3}{c|}{HW-Bench} & \multicolumn{3}{c}{OS-Bench} \\ \cmidrule(l){2-19} 
 & CSR & UR & \multicolumn{1}{c|}{ULR} & CSR & UR & ULR & CSR & UR & \multicolumn{1}{c|}{ULR} & CSR & UR & ULR & CSR & UR & \multicolumn{1}{c|}{ULR} & CSR & UR & ULR \\ \midrule
Instruction & 20.00 & 22.00 & \multicolumn{1}{c|}{6.98} & 6.54 & 28.04 & 10.29 & 20.00 & 25.00 & \multicolumn{1}{c|}{9.14} & 10.28 & 46.73 & 17.70 & 9.00 & 12.00 & \multicolumn{1}{c|}{3.27} & 2.80 & 15.89 & 4.58 \\
ICL & 22.00 & 14.00 & \multicolumn{1}{c|}{2.75} & 24.30 & 19.63 & 7.46 & 26.00 & 35.00 & \multicolumn{1}{c|}{10.02} & 12.15 & 25.23 & 7.62 & 17.00 & 14.00 & \multicolumn{1}{c|}{2.75} & 12.15 & 25.23 & 7.62 \\
RAG & 27.00 & 17.00 & \multicolumn{1}{c|}{5.28} & 13.08 & 25.23 & 10.15 & 36.00 & 38.00 & \multicolumn{1}{c|}{10.53} & 12.15 & 35.51 & 13.22 & 14.00 & 20.00 & \multicolumn{1}{c|}{8.66} & 6.54 & 23.36 & 7.27 \\
Vert & 40.00 & 33.00 & \multicolumn{1}{c|}{11.02} & 14.02 & 26.17 & 8.09 & 35.00 & 45.00 & \multicolumn{1}{c|}{14.59} & 28.97 & 60.75 & 20.76 & 22.00 & 25.00 & \multicolumn{1}{c|}{5.06} & 9.35 & 18.69 & 5.27 \\
IRENE & 39.00 & 33.00 & \multicolumn{1}{c|}{11.38} & 17.76 & 37.38 & 14.90 & 30.00 & 23.00 & \multicolumn{1}{c|}{7.87} & 23.36 & 42.99 & 15.63 & 31.00 & 13.00 & \multicolumn{1}{c|}{4.01} & 15.89 & 21.50 & 5.40 \\
SFT & 49.00 & 11.00 & \multicolumn{1}{c|}{1.88} & 28.04 & \textbf{9.35} & 2.18 & 51.00 & \textbf{10.00} & \multicolumn{1}{c|}{\textbf{2.14}} & \textbf{41.12} & 27.10 & 5.50 & 35.00 & 7.00 & \multicolumn{1}{c|}{1.15} & 13.08 & \textbf{4.67} & 1.67 \\ \midrule
\tool & \textbf{49.00} & \textbf{5.00} & \multicolumn{1}{c|}{\textbf{0.67}} & \textbf{40.19} & 10.28 & \textbf{1.91} & \textbf{54.00} & 19.00 & \multicolumn{1}{c|}{5.69} & 38.32 & \textbf{24.30} & \textbf{4.14} & \textbf{39.00} & \textbf{3.00} & \multicolumn{1}{c|}{\textbf{0.29}} & \textbf{19.63} & 8.41 & \textbf{1.59} \\ \bottomrule
\end{tabular}
\end{table*}
To assess the applicability of \tool in industrial scenarios, we collaborate with Huawei and randomly select 100 C functions from their product-line codebase, denoted as \textbf{HW-Bench}. As 100 functions from a single industrial codebase may not fully reflect this category of code, we additionally collect 107 C functions from the Linux kernel, denoted as \textbf{OS-Bench}. Together, the two benchmarks offer a more representative view of the C code that real-world Rust migration efforts target.
We apply \tool to translate these functions into Rust and compare it against the same baselines used in RQ1. Since no test cases are available for these proprietary or kernel-level functions, the CA metric cannot be computed and we report only CSR, UR, and ULR. The results are shown in Table~\ref{tab:rq3}.

\textbf{Comparison of translation accuracy.}
\tool consistently achieves the highest CSR across most LLMs on both benchmarks. 
For example, Qwen2.5-Coder-3B-Instruct augmented with \tool improves CSR by 16.17\% on HW-Bench and 22.90\% on OS-Bench. 
Across the three models, \tool yields average CSR improvements of 18.28\% and 16.51\%, respectively. 
These results indicate that \tool is effective at producing syntactically valid Rust code for complex industrial programs.

\textbf{Comparison of translation safety.}
In terms of safety, Table~\ref{tab:rq3} shows that \tool achieves the lowest UR and ULR across most LLMs. 
On average, \tool reduces UR and ULR by 13.06\% and 4.37\% on HW-Bench, respectively, 
and by 13.08\% and 6.64\% on OS-Bench. 
These results highlight the superiority of \tool in generating safer Rust code compared to other baselines.

Overall, the results indicate that \tool not only performs reliably in the public benchmark, but also generalizes well to industrial scenario. To further characterize the used benchmarks, following prior work~\cite{FullStack}, we analyze their application-domain distribution (Fig.~\ref{domain}). Both contain a substantially higher proportion of operating-system and system-interaction code than xCodeEval, and OS-Bench in particular consists entirely of OS-level functions. 
Given that the two benchmarks exhibit similar properties, \tool's strong performance on these benchmarks further shows its robustness in handling real-world systems code.
\begin{tcolorbox}
\textbf{Answer to RQ3:} 
\tool improves syntactic validity and safety in industrial scenarios. Across HW-Bench and OS-Bench under three LLMs, \tool achieves an average absolute improvement of 17.40\% in CSR, together with average reductions of 13.07\% in UR and 5.50\% in ULR over baselines. 
\end{tcolorbox}

\section{Discussion}
\label{sec:discussion}

\begin{figure}[t]
    \centering
    \includegraphics[width=\linewidth]{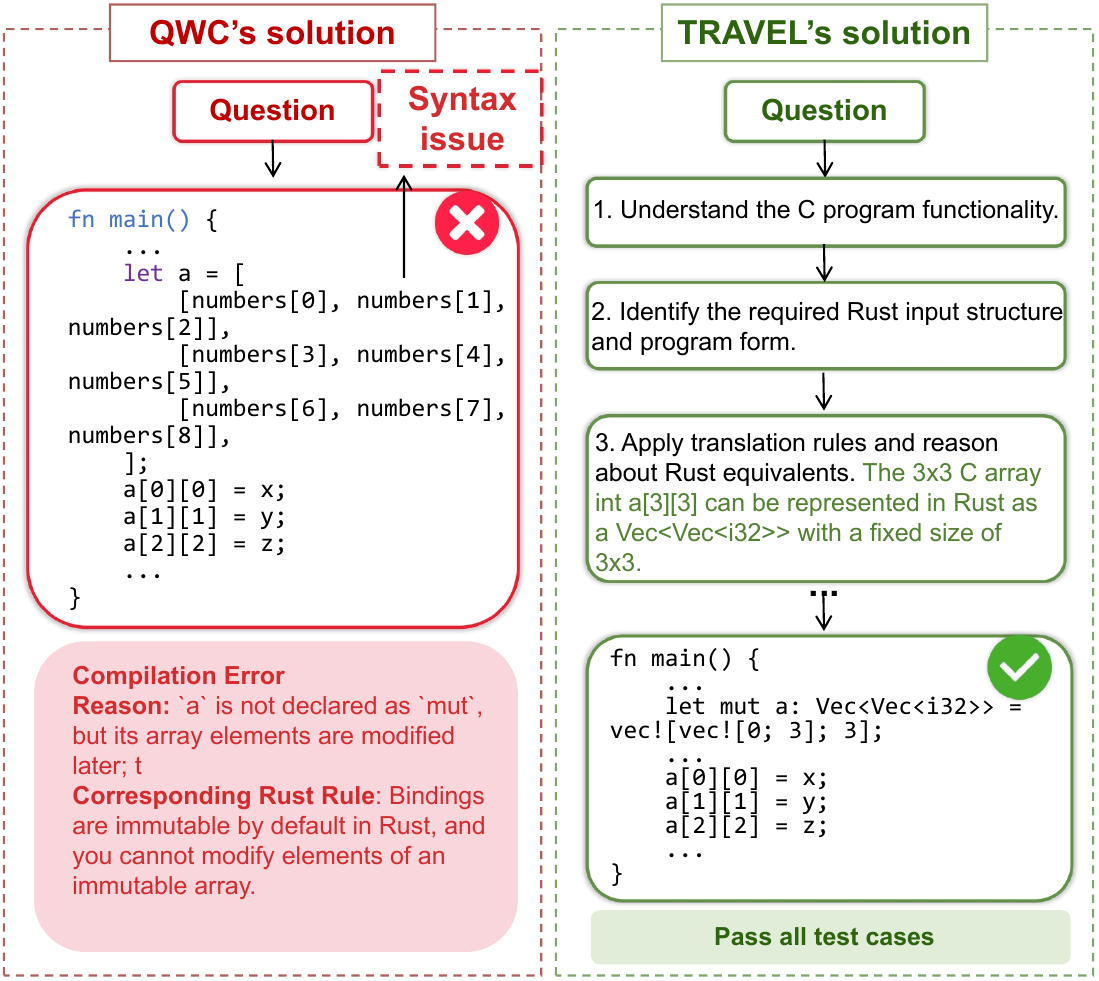}
    \caption{A case demonstrating that Rule-guided MCTS can effectively apply Rust-specific rules to avoid syntax issues. QWC refers to Qwen2.5-Coder-7B-Instruct, which is also used as the backbone of our tool.}
    \label{case1}
\end{figure}

\begin{figure}[t]
    \centering
    \includegraphics[width=\linewidth]{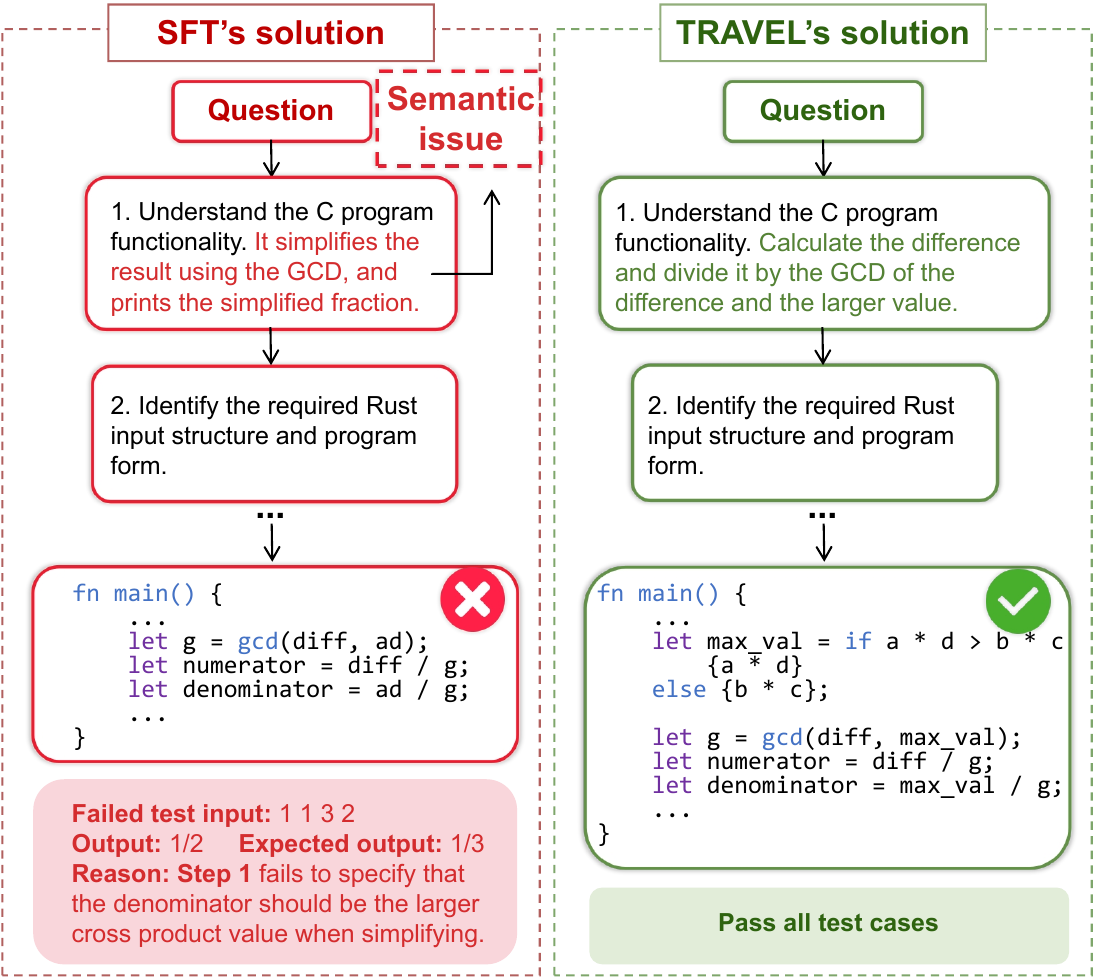}
    \caption{A case showing that dual-reward reinforcement learning further enhances the model’s semantic understanding compared to SFT. All models used here are based on the same backbone, QWC-7B-Instruct.
}
    \label{case2}
\end{figure}
\subsection{Why Does \tool Work?}
\label{subsec:why_works}

In this section, we identify the following two advantages of \tool, which can explain its effectiveness in C-to-Rust translation. For each advantage, we demonstrate the corresponding cases, as shown in Fig.~\ref{case1} and Fig.~\ref{case2}.

\textbf{Advantage 1: Rule-guided MCTS enables the model to handle Rust-specific syntactic rules.}
A common failure of vanilla LLM translation is the violation of Rust-specific constraints (e.g., ownership, borrowing, lifetimes, and FFI signatures) that have no direct counterpart in C.
As shown in Fig.~\ref{case1}, the QWC baseline produces a candidate that looks plausible but breaks Rust's borrow-checker rule, leading to a compilation failure.
In contrast, \tool provides rule-aware guidance during each expansion step via prompt conditioning, which biases the search toward rule-consistent reasoning paths.
As a result, the search converges on a translation that simultaneously respects the language-specific rules and preserves the original behavior.
This mechanism explains why \tool consistently improves CSR across all backbones.

\textbf{Advantage 2: Dual-reward reinforcement learning improves semantic reasoning.} 
SFT provides reasoning supervision but does not ensure that reasoning paths lead to behavior-preserving Rust code. 
As shown on the left of Fig.~\ref{case2}, the SFT-only model produces syntactically valid Rust, but fails to capture a key semantic constraint: the denominator should use the larger cross-product value, resulting in incorrect behavior. 

Our dual-reward RL addresses this by combining a \emph{paths reward}, which favors rule-aligned reasoning steps, and an \emph{execution reward}, which favors translations that pass the test suite. 
These signals guide the model to improve both intermediate reasoning and final outputs. 
As shown on the right of Fig.~\ref{case2}, GRPO enables the same QWC backbone to correctly identify the larger cross-product value, producing behavior-preserving code that passes all tests.

\subsection{Threats to Validity}
\label{subsec:threats}

For our study, we discuss the following threats to validity.

\textbf{Dependence on the underlying LLMs.}
The effectiveness of \tool may be influenced by the capability of the base LLM used for reasoning path generation and translation. 
Stronger models may generate better candidate reasoning paths during MCTS, while weaker models may limit the quality of both search and reinforcement learning. 
To mitigate this threat, we evaluate \tool on multiple representative LLMs rather than relying on a single model. 
The consistent improvements across models suggest that the effectiveness of \tool does not depend on a specific backbone alone.

\textbf{Imperfect verification from LLM-generated test cases.}
Our framework relies on LLM-generated tests to construct Verifiable Components for evaluation and reinforcement learning rewards. 
While this enables scalable semantic verification, such tests may not fully cover corner cases, so translated programs may pass tests yet deviate from the original C code on unseen inputs. 
Prior work~\cite{yang2024} shows that even a 7B model can achieve 87.20\% line coverage with five generated tests, suggesting that LLM-generated tests can provide reasonably strong coverage. 
Given that we use a stronger model (Qwen2.5-Coder-32B-Instruct), we expect comparable or better coverage. 
To mitigate this threat, we generate multiple test inputs per function and ground all outputs by executing the original C program. 
Nevertheless, our evaluation reflects test-based semantic verification rather than full formal equivalence.

\textbf{Potential data leakage.}
As pretraining datasets are not publicly available, we cannot completely rule out potential data leakage. Nevertheless, the limited performance of standard prompting suggests that any such effect is unlikely to be significant. The notable improvements brought by our framework further indicate that the observed gains are attributable to our method rather than simple memorization by the models. Overall, these results support the robustness of our approach and its ability to generalize beyond potential training data overlap.


\section{Related Work}
\label{sec:related_work}

Rule-based techniques apply hand-crafted transformations on top of compiler infrastructures to map C idioms onto Rust counterparts.
A representative example is C2Rust~\cite{immunant2022c2rust}, a Clang/LLVM-based pipeline that produces syntactically equivalent but largely \emph{unsafe} Rust through the C-Rust FFI, and follow-up work extends it along complementary dimensions, including pointer aliasing~\cite{Emre2021,Emre2023}, concurrent lock APIs~\cite{concrat}, ownership inference for pointer-type conversion~\cite{ownership}, and generics and standard-library substitution~\cite{genc2rust,forcrat}.
While interpretable, these approaches rely on expert-defined rules and have limited coverage for real-world programs~\cite{Hong_static}. 

LLM-based methods instead cast translation as a sequence generation task and exploit cross-language patterns learned from large code corpora.
Vert~\cite{VERT} combines few-shot prompting with compiler feedback, while several approaches incorporate program analysis or hybrid pipelines for iterative repair~\cite{C2S,Multi}.
Other methods further focus on rewriting raw pointers and eliminating residual safety defects.~\cite{PR2,SafeTrans}.

In contrast, \tool integrates rule-guided MCTS with reinforcement learning to jointly improve rule compliance and semantic preservation, producing translations that are both compilable and behaviorally faithful.

\section{Conclusion}
\label{sec:conclusion}

In this paper, we propose \tool, an LLM-based framework for C-to-Rust translation. 
\tool improves translation quality by combining rule-guided reasoning-path search with reinforcement learning to enhance both rule compliance and semantic preservation. 
Extensive experiments demonstrate that \tool consistently improves translation accuracy across multiple models and datasets. 
Ablation studies further confirm the effectiveness of each module. 
In future work, we plan to integrate the TRAVEL-trained model into an agent-based migration framework to support repository-level C-to-Rust migration through cross-function reasoning and dependency analysis. We also plan to construct sandbox-based execution environments for larger industrial codebases to enable more scalable and rigorous semantic validation, while extending the framework to support additional programming languages.

\section*{Data Availability Statement}

All data and code are publicly available at~\cite{ourrepo}.



\bibliographystyle{ACM-Reference-Format}
\bibliography{references}

@article{sheng2024hybridflow,
    title   = {HybridFlow: A Flexible and Efficient RLHF Framework},
    author  = {Guangming Sheng and Chi Zhang and Zilingfeng Ye and Xibin Wu and Wang Zhang and Ru Zhang and Yanghua Peng and Haibin Lin and Chuan Wu},
    year    = {2024},
    journal = {arXiv preprint arXiv: 2409.19256}
  }

@misc{openai_gpt5,
  author       = {{OpenAI}},
  title        = {GPT-5.4-mini},
  year         = {2026},
  howpublished = {\url{https://platform.openai.com}},
  note         = {Accessed: 2026-04-30}
}

@software{ourrepo,
  author       = {Luo-feng-hitsz},
  title        = {Luo-feng-hitsz/TRAVEL: v1.0.0},
  month        = jul,
  year         = 2026,
  publisher    = {Zenodo},
  version      = {v1.0.0},
  doi          = {10.5281/zenodo.21132684},
  url          = {https://doi.org/10.5281/zenodo.21132684},
}

@article{crownllm,
  author       = {Hasan Ferit Eniser and
                  Hanliang Zhang and
                  Cristina David and
                  Meng Wang and
                  Maria Christakis and
                  Brandon Paulsen and
                  Joey Dodds and
                  Daniel Kroening},
  title        = {Towards Translating Real-World Code with LLMs: {A} Study of Translating
                  to Rust},
  journal      = {CoRR},
  volume       = {abs/2405.11514},
  year         = {2024},
  url          = {https://doi.org/10.48550/arXiv.2405.11514},
  doi          = {10.48550/ARXIV.2405.11514},
  eprinttype    = {arXiv},
  eprint       = {2405.11514},
  bibsource    = {dblp computer science bibliography, https://dblp.org}
}

@article{Multi,
  author       = {Tianyang Zhou and
                  Haowen Lin and
                  Somesh Jha and
                  Mihai Christodorescu and
                  Kirill Levchenko and
                  Varun Chandrasekaran},
  title        = {LLM-Driven Multi-step Translation from {C} to Rust using Static Analysis},
  journal      = {CoRR},
  volume       = {abs/2503.12511},
  year         = {2025},
  url          = {https://doi.org/10.48550/arXiv.2503.12511},
  doi          = {10.48550/ARXIV.2503.12511},
  eprinttype    = {arXiv},
  eprint       = {2503.12511},
  bibsource    = {dblp computer science bibliography, https://dblp.org}
}

@article{C2S,
  author       = {Vikram Nitin and
                  Rahul Krishna and
                  Luiz Lemos do Valle and
                  Baishakhi Ray},
  title        = {C2SaferRust: Transforming {C} Projects into Safer Rust with NeuroSymbolic
                  Techniques},
  journal      = {CoRR},
  volume       = {abs/2501.14257},
  year         = {2025},
  url          = {https://doi.org/10.48550/arXiv.2501.14257},
  doi          = {10.48550/ARXIV.2501.14257},
  eprinttype    = {arXiv},
  eprint       = {2501.14257},
  bibsource    = {dblp computer science bibliography, https://dblp.org}
}

@article{DBLP:journals/corr/abs-2412-14234,
  author       = {Manish Shetty and
                  Naman Jain and
                  Adwait Godbole and
                  Sanjit A. Seshia and
                  Koushik Sen},
  title        = {Syzygy: Dual Code-Test {C} to (safe) Rust Translation using LLMs and
                  Dynamic Analysis},
  journal      = {CoRR},
  volume       = {abs/2412.14234},
  year         = {2024}
}

@article{DBLP:journals/corr/abs-2501-14257,
  author       = {Vikram Nitin and
                  Rahul Krishna and
                  Luiz Lemos do Valle and
                  Baishakhi Ray},
  title        = {C2SaferRust: Transforming {C} Projects into Safer Rust with NeuroSymbolic
                  Techniques},
  journal      = {CoRR},
  volume       = {abs/2501.14257},
  year         = {2025}
}

@article{DBLP:journals/corr/abs-2409-10506,
  author       = {Momoko Shiraishi and
                  Takahiro Shinagawa},
  title        = {Context-aware Code Segmentation for C-to-Rust Translation using Large
                  Language Models},
  journal      = {CoRR},
  volume       = {abs/2409.10506},
  year         = {2024}
}

@inproceedings{Hong_static,
  author       = {Jaemin Hong},
  title        = {Improving Automatic C-to-Rust Translation with Static Analysis},
  booktitle    = {45th {IEEE/ACM} International Conference on Software Engineering:
                  {ICSE} 2023 Companion Proceedings, Melbourne, Australia, May 14-20,
                  2023},
  pages        = {273--277},
  publisher    = {{IEEE}},
  year         = {2023},
  url          = {https://doi.org/10.1109/ICSE-Companion58688.2023.00074},
  doi          = {10.1109/ICSE-COMPANION58688.2023.00074},
  bibsource    = {dblp computer science bibliography, https://dblp.org}
}

@misc{rustforlinux_nova,
  author       = {{Rust for Linux}},
  title        = {Nova: a Rust-based GPU Driver},
  year         = {2024},
  howpublished = {\url{https://rust-for-linux.com/nova-gpu-driver}},
  note         = {Accessed: 2025-05-19}
}

@misc{cisa_memory_safety_2023,
  author       = {{Cybersecurity and Infrastructure Security Agency (CISA)}},
  title        = {The Urgent Need for Memory Safety in Software Products},
  year         = {2023},
  howpublished = {\url{https://www.cisa.gov/news-events/news/urgent-need-memory-safety-software-products}},
  note         = {Accessed: 2026-04-30}
}

@misc{project_zero_memory_safety,
  author       = {{Google Project Zero}},
  title        = {Memory Safety and the Prevalence of Exploitable Vulnerabilities},
  year         = {2021},
  howpublished = {\url{https://googleprojectzero.blogspot.com/2021/02/a-year-of-windows-kernel-exploitation.html}},
  note         = {Accessed: 2026-04-30}
}

@article{FullStack,
  author       = {Yao Cheng and
                  Jianfeng Chen and
                  Jie Chen and
                  Li Chen and
                  Liyu Chen and
                  Wentao Chen and
                  Zhengyu Chen and
                  Shijie Geng and
                  Aoyan Li and
                  Bo Li and
                  Bowen Li and
                  Linyi Li and
                  Boyi Liu and
                  Jerry Liu and
                  Kaibo Liu and
                  Qi Liu and
                  Shukai Liu and
                  Siyao Liu and
                  Tianyi Liu and
                  Tingkai Liu and
                  Yongfei Liu and
                  Rui Long and
                  Jing Mai and
                  Guanghan Ning and
                  Z. Y. Peng and
                  Kai Shen and
                  Jiahao Su and
                  Jing Su and
                  Tao Sun and
                  Yifan Sun and
                  Yunzhe Tao and
                  Guoyin Wang and
                  Siwei Wang and
                  Xuwu Wang and
                  Yite Wang and
                  Zihan Wang and
                  Jinxiang Xia and
                  Liang Xiang and
                  Xia Xiao and
                  Yongsheng Xiao and
                  Chenguang Xi and
                  Shulin Xin and
                  Jingjing Xu and
                  Shikun Xu and
                  Hongxia Yang and
                  Jack Yang and
                  Yingxiang Yang and
                  Jianbo Yuan and
                  Jun Zhang and
                  Yufeng Zhang and
                  Yuyu Zhang and
                  Shen Zheng and
                  He Zhu and
                  Ming Zhu},
  title        = {FullStack Bench: Evaluating LLMs as Full Stack Coders},
  journal      = {CoRR},
  volume       = {abs/2412.00535},
  year         = {2024},
  url          = {https://doi.org/10.48550/arXiv.2412.00535},
  doi          = {10.48550/ARXIV.2412.00535},
  eprinttype    = {arXiv},
  eprint       = {2412.00535},
  bibsource    = {dblp computer science bibliography, https://dblp.org}
}

@article{2023claude,
  title   = {Introducing Claude},
  author  = {{Anthropic}},
  journal = {Anthropic Blog},
  year    = {2023},
  howpublished = {\url{https://www.anthropic.com/index/introducing-claude}},
  note    = {Accessed: 2026-04-27}
}

@article{openai2024gpt4technicalreport,
  title={Gpt-4 technical report},
  author={Achiam, Josh and Adler, Steven and Agarwal, Sandhini and Ahmad, Lama and Akkaya, Ilge and Aleman, Florencia Leoni and Almeida, Diogo and Altenschmidt, Janko and Altman, Sam and Anadkat, Shyamal and others},
  journal={arXiv preprint arXiv:2303.08774},
  year={2023}
}

@article{hu2022lora,
  title={Lora: Low-rank adaptation of large language models.},
  author={Hu, Edward J and Shen, Yelong and Wallis, Phillip and Allen-Zhu, Zeyuan and Li, Yuanzhi and Wang, Shean and Wang, Liang and Chen, Weizhu and others},
  journal={Iclr},
  volume={1},
  number={2},
  pages={3},
  year={2022}
}

@article{yang2024,
  author       = {Zhen Yang and
                  Fang Liu and
                  Zhongxing Yu and
                  Jacky Wai Keung and
                  Jia Li and
                  Shuo Liu and
                  Yifan Hong and
                  Xiaoxue Ma and
                  Zhi Jin and
                  Ge Li},
  title        = {Exploring and Unleashing the Power of Large Language Models in Automated
                  Code Translation},
  journal      = {Proc. {ACM} Softw. Eng.},
  volume       = {1},
  number       = {{FSE}},
  pages        = {1585--1608},
  year         = {2024},
  url          = {https://doi.org/10.1145/3660778},
  doi          = {10.1145/3660778},
  bibsource    = {dblp computer science bibliography, https://dblp.org}
}

@inproceedings{DBLP:conf/cav/ZhangDYW23,
  author       = {Hanliang Zhang and
                  Cristina David and
                  Yijun Yu and
                  Meng Wang},
  title        = {Ownership Guided {C} to Rust Translation},
  booktitle    = {{CAV} {(3)}},
  series       = {Lecture Notes in Computer Science},
  volume       = {13966},
  pages        = {459--482},
  publisher    = {Springer},
  year         = {2023}
}

@inproceedings{DBLP:conf/kbse/HongR24,
  author       = {Jaemin Hong and
                  Sukyoung Ryu},
  title        = {To Tag, or Not to Tag: Translating C's Unions to Rust's Tagged Unions},
  booktitle    = {{ASE}},
  pages        = {40--52},
  publisher    = {{ACM}},
  year         = {2024}
}

@inproceedings{DBLP:conf/icse/LingYWWCH22,
  author       = {Michael Ling and
                  Yijun Yu and
                  Haitao Wu and
                  Yuan Wang and
                  James R. Cordy and
                  Ahmed E. Hassan},
  title        = {In Rust We Trust - {A} Transpiler from Unsafe {C} to Safer Rust},
  booktitle    = {ICSE-Companion},
  pages        = {354--355},
  publisher    = {{ACM/IEEE}},
  year         = {2022}
}

@inproceedings{IRENE,
  author    = {Luo, Feng and Ji, Kexing and Gao, Cuiyun and Gao, Shuzheng and Feng, Jia and Liu, Kui and Xia, Xin and Lyu, Michael},
  title     = {Integrating Rules and Semantics for {LLM}-Based {C}-to-{R}ust Translation},
  booktitle = {Proceedings of the 41st IEEE International Conference on Software Maintenance and Evolution (ICSME)},
  pages     = {685--696},
  year      = {2025},
  month     = {9},
  doi       = {10.1109/ICSME64153.2025.00069}
}

@misc{miller2019trends,
  author       = {Cimpanu, Catalin},
  title        = {{Microsoft}: 70 percent of all security bugs are memory safety issues},
  howpublished = {ZDNet},
  year         = {2019},
  month        = feb,
  day          = {11},
  note         = {\url{https://www.zdnet.com/article/microsoft-70-percent-of-all-security-bugs-are-memory-safety-issues/}}
}

@misc{googleandroid2022memorysafe,
  author       = {Vander Stoep, Jeffrey and Hines, Stephen},
  title        = {Memory Safe Languages in {Android} 13},
  howpublished = {Google Security Blog},
  year         = {2022},
  month        = dec,
  day          = {1},
  note         = {\url{https://security.googleblog.com/2022/12/memory-safe-languages-in-android-13.html}}
}

@inproceedings{agache2020firecracker,
  author    = {Agache, Alexandru and Brooker, Marc and Florescu, Andreea and Iordache, Alexandra and Liguori, Anthony and Neugebauer, Rolf and Piwonka, Phil and Popa, Diana-Maria},
  title     = {Firecracker: Lightweight Virtualization for Serverless Applications},
  booktitle = {17th {USENIX} Symposium on Networked Systems Design and Implementation ({NSDI} 20)},
  pages     = {419--434},
  year      = {2020},
  publisher = {{USENIX} Association},
  isbn      = {978-1-939133-13-7},
  url       = {https://www.usenix.org/conference/nsdi20/presentation/agache}
}

@misc{microsoft2023rustwindows,
  author       = {Claburn, Thomas},
  title        = {{Microsoft} is busy rewriting core {Windows} library code in memory-safe {Rust}},
  howpublished = {The Register},
  year         = {2023},
  month        = apr,
  day          = {27},
  note         = {\url{https://www.theregister.com/2023/04/27/microsoft_windows_rust/}}
}

@misc{immunant2022c2rust,
  author       = {Immunant},
  title        = {C2Rust},
  year         = {2022},
  howpublished = {\url{https://github.com/immunant/c2rust}}
}

@misc{clang,
  author = {{LLVM Project}},
  title = {Clang: A C Language Family Frontend for LLVM},
  howpublished = {\url{https://clang.llvm.org/}},
  note = {Accessed: 2025-05-26}
}

@inproceedings{DBLP:conf/cgo/LattnerA04,
  author       = {Chris Lattner and
                  Vikram S. Adve},
  title        = {{LLVM:} {A} Compilation Framework for Lifelong Program Analysis {\&}
                  Transformation},
  booktitle    = {2nd {IEEE} / {ACM} International Symposium on Code Generation and
                  Optimization {(CGO} 2004), 20-24 March 2004, San Jose, CA, {USA}},
  pages        = {75--88},
  publisher    = {{IEEE} Computer Society},
  year         = {2004},
  url          = {https://doi.org/10.1109/CGO.2004.1281665},
  doi          = {10.1109/CGO.2004.1281665},
  bibsource    = {dblp computer science bibliography, https://dblp.org}
}

@inproceedings{concrat,
  author       = {Jaemin Hong and
                  Sukyoung Ryu},
  title        = {Concrat: An Automatic C-to-Rust Lock {API} Translator for Concurrent
                  Programs},
  booktitle    = {45th {IEEE/ACM} International Conference on Software Engineering,
                  {ICSE} 2023, Melbourne, Australia, May 14-20, 2023},
  pages        = {716--728},
  publisher    = {{IEEE}},
  year         = {2023},
  url          = {https://doi.org/10.1109/ICSE48619.2023.00069},
  doi          = {10.1109/ICSE48619.2023.00069},
  bibsource    = {dblp computer science bibliography, https://dblp.org}
}

@inproceedings{ownership,
  author       = {Hanliang Zhang and
                  Cristina David and
                  Yijun Yu and
                  Meng Wang},
  editor       = {Constantin Enea and
                  Akash Lal},
  title        = {Ownership Guided {C} to Rust Translation},
  booktitle    = {Computer Aided Verification - 35th International Conference, {CAV}
                  2023, Paris, France, July 17-22, 2023, Proceedings, Part {III}},
  series       = {Lecture Notes in Computer Science},
  volume       = {13966},
  pages        = {459--482},
  publisher    = {Springer},
  year         = {2023},
  url          = {https://doi.org/10.1007/978-3-031-37709-9\_22},
  doi          = {10.1007/978-3-031-37709-9\_22},
  bibsource    = {dblp computer science bibliography, https://dblp.org}
}

@article{Emre2021,
  author       = {Mehmet Emre and
                  Ryan Schroeder and
                  Kyle Dewey and
                  Ben Hardekopf},
  title        = {Translating {C} to safer Rust},
  journal      = {Proc. {ACM} Program. Lang.},
  volume       = {5},
  number       = {{OOPSLA}},
  pages        = {1--29},
  year         = {2021},
  url          = {https://doi.org/10.1145/3485498},
  doi          = {10.1145/3485498},
  bibsource    = {dblp computer science bibliography, https://dblp.org}
}

@article{Emre2023,
  author       = {Mehmet Emre and
                  Peter Boyland and
                  Aesha Parekh and
                  Ryan Schroeder and
                  Kyle Dewey and
                  Ben Hardekopf},
  title        = {Aliasing Limits on Translating {C} to Safe Rust},
  journal      = {Proc. {ACM} Program. Lang.},
  volume       = {7},
  number       = {{OOPSLA1}},
  pages        = {551--579},
  year         = {2023},
  url          = {https://doi.org/10.1145/3586046},
  doi          = {10.1145/3586046},
  bibsource    = {dblp computer science bibliography, https://dblp.org}
}

@INPROCEEDINGS {genc2rust,
author = { Wu, Xiafa and Demsky, Brian },
booktitle = { 2025 IEEE/ACM 47th International Conference on Software Engineering (ICSE) },
title = {{ GenC2Rust: Towards Generating Generic Rust Code from C }},
year = {2025},
volume = {},
ISSN = {},
pages = {90-102},
doi = {10.1109/ICSE55347.2025.00127},
url = {https://doi.ieeecomputersociety.org/10.1109/ICSE55347.2025.00127},
publisher = {IEEE Computer Society},
address = {Los Alamitos, CA, USA},
month =May}

@inproceedings{forcrat,
author = {Hong, Jaemin and Ryu, Sukyoung},
title = {Forcrat: Automatic I/O API Translation from C to Rust via Origin and Capability Analysis},
year = {2025},
publisher = {IEEE Press},
url = {https://doi.org/10.1109/ASE63991.2025.00130},
doi = {10.1109/ASE63991.2025.00130},
booktitle = {2025 40th IEEE/ACM International Conference on Automated Software Engineering (ASE)},
pages = {1541–1552},
numpages = {12},
location = {Seoul, Korea, Republic of}
}

@article{PR2,
  title={Pr2: Peephole raw pointer rewriting with llms for translating c to safer rust},
  author={Gao, Yifei and Wang, Chengpeng and Huang, Pengxiang and Liu, Xuwei and Zheng, Mingwei and Zhang, Xiangyu},
  journal={arXiv preprint arXiv:2505.04852},
  year={2025}
}

@article{SafeTrans,
  title={Safetrans: Llm-assisted transpilation from c to rust},
  author={Farrukh, Muhammad and Shah, Smeet and Coskun, Baris and Polychronakis, Michalis},
  journal={arXiv preprint arXiv:2505.10708},
  year={2025}
}

@article{codegemma,
  title={Codegemma: Open code models based on gemma},
  author={Team, CodeGemma and Zhao, Heri and Hui, Jeffrey and Howland, Joshua and Nguyen, Nam and Zuo, Siqi and Hu, Andrea and Choquette-Choo, Christopher A and Shen, Jingyue and Kelley, Joe and others},
  journal={arXiv preprint arXiv:2406.11409},
  year={2024}
}

@inproceedings{coulom2006efficient,
  title={Efficient selectivity and backup operators in Monte-Carlo tree search},
  author={Coulom, R{\'e}mi},
  booktitle={International conference on computers and games},
  pages={72--83},
  year={2006},
  organization={Springer}
}

@article{silver2016mastering,
  title={Mastering the game of Go with deep neural networks and tree search},
  author={Silver, David and Huang, Aja and Maddison, Chris J and Guez, Arthur and Sifre, Laurent and Van Den Driessche, George and Schrittwieser, Julian and Antonoglou, Ioannis and Panneershelvam, Veda and Lanctot, Marc and others},
  journal={nature},
  volume={529},
  number={7587},
  pages={484--489},
  year={2016},
  publisher={Nature Publishing Group UK London}
}

@article{silver2017mastering,
  title={Mastering the game of go without human knowledge},
  author={Silver, David and Schrittwieser, Julian and Simonyan, Karen and Antonoglou, Ioannis and Huang, Aja and Guez, Arthur and Hubert, Thomas and Baker, Lucas and Lai, Matthew and Bolton, Adrian and others},
  journal={nature},
  volume={550},
  number={7676},
  pages={354--359},
  year={2017},
  publisher={Nature Publishing Group UK London}
}

@article{PUCT,
  title={Multi-armed bandits with episode context},
  author={Rosin, Christopher D},
  journal={Annals of Mathematics and Artificial Intelligence},
  volume={61},
  number={3},
  pages={203--230},
  year={2011},
  publisher={Springer}
}

@article{mahan2025generative,
  title={Generative reward models},
  author={Mahan, Dakota and Van Phung, Duy and Rafailov, Rafael and Blagden, Chase and Lile, Nathan and Castricato, Louis and Fr{\"a}nken, Jan-Philipp and Finn, Chelsea and Albalak, Alon},
  journal={arXiv preprint arXiv:2410.12832},
  year={2024}
}

@article{akhauri2025performance,
  title={Performance prediction for large systems via text-to-text regression},
  author={Akhauri, Yash and Lewandowski, Bryan and Lin, Cheng-Hsi and Reyes, Adrian N and Forbes, Grant C and Wongpanich, Arissa and Yang, Bangding and Abdelfattah, Mohamed S and Perel, Sagi and Song, Xingyou},
  journal={arXiv preprint arXiv:2506.21718},
  year={2025}
}

@article{guo2025deepseek,
  title={DeepSeek-R1 incentivizes reasoning in LLMs through reinforcement learning},
  author={Guo, Daya and Yang, Dejian and Zhang, Haowei and Song, Junxiao and Wang, Peiyi and Zhu, Qihao and Xu, Runxin and Zhang, Ruoyu and Ma, Shirong and Bi, Xiao and others},
  journal={Nature},
  volume={645},
  number={8081},
  pages={633--638},
  year={2025},
  publisher={Nature Publishing Group UK London}
}

@inproceedings{xcode,
  author       = {Mohammad Abdullah Matin Khan and
                  M. Saiful Bari and
                  Xuan Do Long and
                  Weishi Wang and
                  Md. Rizwan Parvez and
                  Shafiq Joty},
  editor       = {Lun{-}Wei Ku and
                  Andre Martins and
                  Vivek Srikumar},
  title        = {XCodeEval: An Execution-based Large Scale Multilingual Multitask Benchmark
                  for Code Understanding, Generation, Translation and Retrieval},
  booktitle    = {Proceedings of the 62nd Annual Meeting of the Association for Computational
                  Linguistics (Volume 1: Long Papers), {ACL} 2024, Bangkok, Thailand,
                  August 11-16, 2024},
  pages        = {6766--6805},
  publisher    = {Association for Computational Linguistics},
  year         = {2024},
  url          = {https://doi.org/10.18653/v1/2024.acl-long.367},
  doi          = {10.18653/V1/2024.ACL-LONG.367},
  bibsource    = {dblp computer science bibliography, https://dblp.org}
}

@misc{github_os_data,
  author       = {{Torvalds, Linus and Linux kernel contributors}},
  title        = {{Linux kernel source tree}},
  year         = {2026},
  howpublished = {\url{https://github.com/torvalds/linux}},
  note         = {Accessed: 2026-04-27}
}

@inproceedings{ICL,
  author       = {Tom B. Brown and
                  Benjamin Mann and
                  Nick Ryder and
                  Melanie Subbiah and
                  Jared Kaplan and
                  Prafulla Dhariwal and
                  Arvind Neelakantan and
                  Pranav Shyam and
                  Girish Sastry and
                  Amanda Askell and
                  Sandhini Agarwal and
                  Ariel Herbert{-}Voss and
                  Gretchen Krueger and
                  Tom Henighan and
                  Rewon Child and
                  Aditya Ramesh and
                  Daniel M. Ziegler and
                  Jeffrey Wu and
                  Clemens Winter and
                  Christopher Hesse and
                  Mark Chen and
                  Eric Sigler and
                  Mateusz Litwin and
                  Scott Gray and
                  Benjamin Chess and
                  Jack Clark and
                  Christopher Berner and
                  Sam McCandlish and
                  Alec Radford and
                  Ilya Sutskever and
                  Dario Amodei},
  title        = {Language Models are Few-Shot Learners},
  booktitle    = {NeurIPS},
  year         = {2020}
}

@inproceedings{RAG,
  author       = {Wenqi Fan and
                  Yujuan Ding and
                  Liangbo Ning and
                  Shijie Wang and
                  Hengyun Li and
                  Dawei Yin and
                  Tat{-}Seng Chua and
                  Qing Li},
  title        = {A Survey on {RAG} Meeting LLMs: Towards Retrieval-Augmented Large
                  Language Models},
  booktitle    = {{KDD}},
  pages        = {6491--6501},
  publisher    = {{ACM}},
  year         = {2024}
}

@article{VERT,
  title={Vert: Verified equivalent rust transpilation with large language models as few-shot learners},
  author={Yang, Aidan ZH and Takashima, Yoshiki and Paulsen, Brandon and Dodds, Josiah and Kroening, Daniel},
  journal={arXiv preprint arXiv:2404.18852},
  year={2024}
}

@article{SFT,
author = {Yang, Guang and Zhou, Yu and Chen, Xiang and Zhang, Xiangyu and Zhuo, Terry Yue and Chen, Taolue},
title = {Chain-of-Thought in Neural Code Generation: From and for Lightweight Language Models},
year = {2024},
issue_date = {Sept. 2024},
publisher = {IEEE Press},
volume = {50},
number = {9},
issn = {0098-5589},
url = {https://doi.org/10.1109/TSE.2024.3440503},
doi = {10.1109/TSE.2024.3440503},
journal = {IEEE Trans. Softw. Eng.},
month = sep,
pages = {2437–2457},
numpages = {21}
}

@article{hui2024qwen2,
  title   = {Qwen2.5-Coder Technical Report},
  author  = {Hui, Binyuan and Yang, Jian and Cui, Zeyu and Yang, Jiaxi and Liu, Dayiheng and Zhang, Lei and Liu, Tianyu and Zhang, Jiajun and Yu, Bowen and Lu, Keming and Dang, Kai and Fan, Yang and Zhang, Yichang and Yang, An and Men, Rui and Huang, Fei and Zheng, Bo and Miao, Yibo and Quan, Shanghaoran and Feng, Yunlong and Ren, Xingzhang and Ren, Xuancheng and Zhou, Jingren and Lin, Junyang},
  journal = {arXiv preprint arXiv:2409.12186},
  year    = {2024}
}

@article{yang2025qwen3,
  title={Qwen3 technical report},
  author={Yang, An and Li, Anfeng and Yang, Baosong and Zhang, Beichen and Hui, Binyuan and Zheng, Bo and Yu, Bowen and Gao, Chang and Huang, Chengen and Lv, Chenxu and others},
  journal={arXiv preprint arXiv:2505.09388},
  year={2025}
}

@inproceedings{wolf2020transformers,
  title     = {Transformers: State-of-the-Art Natural Language Processing},
  author    = {Wolf, Thomas and Debut, Lysandre and Sanh, Victor and Chaumond, Julien and Delangue, Clement and Moi, Anthony and Cistac, Pierric and Rault, Timothee and Louf, Remi and Funtowicz, Morgan and Davison, Joe and Shleifer, Sam and von Platen, Patrick and Ma, Clara and Jernite, Yacine and Plu, Julien and Xu, Canwen and Le Scao, Teven and Gugger, Sylvain and Drame, Mariama and Lhoest, Quentin and Rush, Alexander M.},
  booktitle = {Proceedings of the 2020 Conference on Empirical Methods in Natural Language Processing: System Demonstrations},
  pages     = {38--45},
  year      = {2020}
}

@inproceedings{paszke2019pytorch,
  title     = {PyTorch: An Imperative Style, High-Performance Deep Learning Library},
  author    = {Paszke, Adam and Gross, Sam and Massa, Francisco and Lerer, Adam and Bradbury, James and Chanan, Gregory and Killeen, Trevor and Lin, Zeming and Gimelshein, Natalia and Antiga, Luca and Desmaison, Alban and Kopf, Andreas and Yang, Edward and DeVito, Zachary and Raison, Martin and Tejani, Alykhan and Chilamkurthy, Sasank and Steiner, Benoit and Fang, Lu and Bai, Junjie and Chintala, Soumith},
  booktitle = {Advances in Neural Information Processing Systems},
  volume    = {32},
  year      = {2019}
}

@inproceedings{DBLP:conf/kbse/GaoWGWZL23,
  author       = {Shuzheng Gao and
                  Xin{-}Cheng Wen and
                  Cuiyun Gao and
                  Wenxuan Wang and
                  Hongyu Zhang and
                  Michael R. Lyu},
  title        = {What Makes Good In-Context Demonstrations for Code Intelligence Tasks
                  with LLMs?},
  booktitle    = {38th {IEEE/ACM} International Conference on Automated Software Engineering,
                  {ASE} 2023, Luxembourg, September 11-15, 2023},
  pages        = {761--773},
  publisher    = {{IEEE}},
  year         = {2023},
  url          = {https://doi.org/10.1109/ASE56229.2023.00109},
  doi          = {10.1109/ASE56229.2023.00109},
  bibsource    = {dblp computer science bibliography, https://dblp.org}
}
\end{document}